\documentclass[aps,pra,twocolumn,showpacs,superscriptaddress]{revtex4}
\usepackage{graphicx}
\usepackage{amsmath}
\usepackage{amssymb}

\input{epsf}
\begin{document}

\title{Hyperspherical explicitly correlated Gaussian approach
for 
few-body systems
with finite angular momentum}

\author{D. Rakshit}
\affiliation{Department of Physics and Astronomy,
Washington State University,
  Pullman, Washington 99164-2814, USA}
\author{D. Blume}
\affiliation{Department of Physics and Astronomy,
Washington State University,
  Pullman, Washington 99164-2814, USA}
\affiliation{ITAMP, Harvard-Smithsonian Center for Astrophysics,
60 Garden Street, Cambridge, Massachusetts 02138, USA}

\date{\today}

\begin{abstract}
Within the hyperspherical framework, the 
solution of the time-independent Schr\"odinger equation
for a $n$-particle system is divided
into two steps, the solution of a Schr\"odinger like
equation in the hyperangular degrees of freedom
and the solution of a set of coupled
Schr\"odinger like hyperradial 
equations. The solutions to the former 
provide effective potentials and coupling matrix elements
that enter 
into the latter set of equations. 
This paper develops a theoretical
framework 
to determine the effective
potentials, as well as the associated coupling matrix
elements, 
for 
few-body systems with
finite angular momentum $L=1$ and negative and positive parity $\Pi$.
The hyperangular channel functions
are expanded in terms of explicitly correlated Gaussian basis
functions and relatively compact expressions
for the matrix elements are derived.
The developed formalism is applicable to any $n$;
however, for $n \ge 6$, the computational
demands are likely beyond present-day 
computational capabilities.
A number of 
calculations relevant to cold atom physics are presented, demonstrating that
the developed approach provides a computationally
efficient means to solving four-body bound and scattering problems
with finite angular momentum
on 
powerful
desktop computers.
Details regarding the implementation are discussed.
\end{abstract}


\maketitle

\section{Introduction}
\label{sec_introduction}
Few-body phenomena play important roles 
across all disciplines of physics,
including
atomic and molecular physics, chemical physics,
nuclear and particle physics, and condensed matter physics.
Progress in solving and understanding the quantum mechanical 
few-body problem has been driven, roughly speaking,
by one or more of the following three aspects:
{\em{(i)}} Using well-established
algorithms, the steady increase of computational resources
has made it possible to tackle problems
that were impossible to tackle a decade or even just a few years ago.
{\em{(ii)}} A number of model systems have been investigated
analytically, semi-analytically or numerically, providing crucial insights
into some of the low-energy processes that govern the few-body
dynamics.
{\em{(iii)}} More efficient numerical schemes that are not only
applicable to the three-body problem
but also to four- and higher-body problems have been developed.

This paper 
extends the 
correlated Gaussian hyperspherical
(CGHS)
or
hyperspherical explicitly
correlated Gaussian (HECG) 
approach~\cite{vonstecherPRA,vonstecherThesis,ritt11}.
In earlier work, von Stecher and 
Greene~\cite{vonstecherPRA,vonstecherThesis,ritt11}
considered three- and four-body systems
with vanishing angular momentum $L$ and positive parity $\Pi$.
Here, we extend the approach
to systems with 
finite angular momentum $L$.
Although the overall scheme developed for systems 
with $L^{\Pi}=0^+$ symmetry carries over to systems with finite angular
momentum, the determination of compact expressions 
for the matrix elements associated
with finite angular momentum states is significantly more involved
than that for states with $L^{\Pi}=0^+$ symmetry.

The HECG approach provides an
efficient numerical scheme for solving few-body problems.
It is a basis set expansion 
type approach, which combines elements of the aspects 
{\em{(i)}}-{\em{(iii)}}
mentioned above.
In particular, the use
of hyperspherical 
coordinates~\cite{delves1,delves2,ballot,macek,shertzer,fano,smith,aquilanti,pack,launay,ritt11,averybook,linreview}
within the framework of explicitly correlated Gaussian (ECG)
basis functions~\cite{cgbook}
allows us to take advantage of
the machinery developed
for bound state calculations while  at the same time
enables us to describe the scattering continuum.
Although significant progress has been 
made~\cite{vonstecherPRA,ritt11,deltuvascatt,suzukiscatt,navratilscatt,orlandiniscatt},
in general, the determination of scattering 
quantities is significantly more involved than that
of bound state quantities, and the four-, five- and higher-body scattering
continua are comparatively poorly understood.
Thus, the framework developed in this work for systems with 
$1^-$ and $1^+$ symmetry
provides a promising step forward.

The HECG approach is quite general and applicable
to a wide range of few-body systems.
As an application, we consider the four-particle system
consisting of three identical fermions and an impurity
whose mass is lighter than that of the 
majority species.
We assume interspecies short-range $s$-wave interactions and
investigate the system properties 
of the energetically lowest-lying
$1^+$ state as a function of the mass ratio $\kappa$
between the majority particles and the impurity particle.
These finite angular momentum states
are interesting since 
universal
four-body bound states have been predicted to exist
if the two-body $s$-wave scattering length is 
positive and $\kappa \gtrsim 9.5$~\cite{blum12}.
Moreover, for $13.38 \lesssim \kappa \lesssim 13.61$,
the $(3,1)$ system with $1^+$ symmetry has been
predicted to support 
four-body Efimov states~\cite{castin};
in this mass ratio regime, three-body Efimov states are
absent~\cite{efimov1,efimov2,braatenReview,petrov}.
This work determines 
and interprets the hyperangular eigen value of the
$(3,1)$ system with infinitely large
interspecies $s$-wave scattering length 
in the limit that the
hyperradius is much larger than the range of the
underlying two-body potential.

The remainder of this paper is organized as follows.
Section~\ref{sec_system} introduces the 
system Hamiltonian and the hyperspherical framework,
while Sec.~\ref{sec_basisfunctions} introduces the 
ECG basis functions used to expand the hyperangular channel functions.
Section~\ref{sec_matrixelements}
discusses the
matrix elements,
applicable to any $n$, needed to calculate the 
effective hyperradial potential curves
and associated coupling matrix elements.
Details regarding the numerical implementation of the
HECG approach and a set of proof-of-principle
calculations for the four-
and five-particle system
are discussed in Sec.~\ref{sec_implementation}.
Section~\ref{sec_applications} applies the
HECG framework to the $(3,1)$ system with $1^+$
symmetry and diverging interspecies $s$-wave
scattering length for various mass ratios $\kappa$.
Lastly, Sec.~\ref{sec_conclusions} summarizes and concludes.
Details regarding the derivation of and final results for the 
fixed hyperradius matrix elements 
are presented in three appendices.
Appendix~\ref{appendix_symmetry}
defines a number of auxiliary quantities that 
depend on the symmetry considered.
Appendix~\ref{appendix_derivation}
outlines exemplarily how to derive the matrix elements
for the three-body system with $1^-$ symmetry.
Appendix~\ref{appendix_general}
summarizes our expressions for a number of quantities that
enter into the final equations for the matrix elements;
these equations apply to all symmetries considered in
this paper.

\section{System Hamiltonian and hyperspherical framework}
\label{sec_system}
We consider an $n$-particle system 
with position vectors $\vec{r}_j$ 
described by the Hamiltonian $H$,
\begin{eqnarray}
\label{eq_ham}
H = \sum_{j=1}^{n} -\frac{\hbar^2}{2m_j} \nabla^2_{\vec{r}_j}
+ V_{\rm{int}},
\end{eqnarray}
where $m_j$ denotes the mass of the $j$$^{\rm{th}}$
particle.
The interaction potential
$V_{\rm{int}}$ is written  as a sum of two-body
potentials $V_{jk}(\vec{r}_{jk})$,
\begin{eqnarray}
V_{\rm{int}} =
\sum_{j<k}^n V_{jk}(\vec{r}_{jk}),
\end{eqnarray}
where $\vec{r}_{jk} = \vec{r}_j - \vec{r}_k$
($r_{jk}=|\vec{r}_{jk}|$).
To
separate off the center of mass degrees of freedom,
we define $n$ 
mass-scaled 
Jacobi vectors
$\vec{\rho}_j$,
\begin{eqnarray}
\vec{\rho}_j = \sum_{k=1}^n  T_{jk} \vec{r}_k.
\end{eqnarray}
The elements $T_{jk}$ form a $n \times n$ matrix.
The explicit forms for $n=3$ and $n=4$
read
\begin{eqnarray}
\underline{T}_{n=3} = 
\left( \begin{array}{ccc}
\sqrt{\mu_1}         & -\sqrt{\mu_1}        & 0 \\
\frac{\sqrt{\mu_2}m_1}{m_1+m_2} & \frac{\sqrt{\mu_2}m_2}{m_1+m_2} & -\sqrt{\mu_2} \\
m_1/\sqrt{\mu_3} & m_2/\sqrt{\mu_3} & m_3/\sqrt{\mu_3}
\end{array}
\right)
\end{eqnarray}
and
\begin{eqnarray}
\underline{T}_{n=4} = \nonumber \\
\left( \begin{array}{cccc}
\sqrt{\mu_1}         & -\sqrt{\mu_1}        & 0         & 0 \\
\frac{\sqrt{\mu_2}m_1}{m_1+m_2} & \frac{\sqrt{\mu_2}m_2}{m_1+m_2} & -\sqrt{\mu_2} & 0 \\
\frac{\sqrt{\mu_3}m_1}{m_1+m_2+m_3} & \frac{\sqrt{\mu_3}m_2}{m_1+m_2+m_3} & \frac{\sqrt{\mu_3}m_3}{m_1+m_2+m_3} & -\sqrt{\mu_3} \\
\frac{m_1}{\sqrt{\mu_4}} & \frac{m_2}{\sqrt{\mu_4}} & \frac{m_3}{\sqrt{\mu_4}} & \frac{m_4}{\sqrt{\mu_4}}
\end{array}
\right),
\end{eqnarray}
where $\mu_j $ denotes the mass associated with the
$j$$^{\rm{th}}$ Jacobi vector,
\begin{eqnarray}
\mu_j = \frac{ \left( \sum_{k=1}^j m_k \right) m_{j+1}}{\sum_{k=1}^{j+1}m_k} 
\mbox{ for } j = 1,\cdots,n-1
\end{eqnarray}
and
\begin{eqnarray}
\mu_n = \sum_{k=1}^n m_k.
\end{eqnarray}
The generalization to $n \ge 5$ is straightforward.
By definition, the $n$$^{\rm{th}}$ Jacobi vector coincides with the 
``mass-scaled'' center
of mass vector of the $n$-particle system.
Although the mass-scaling is not needed to separate off the center of mass
motion, the use of mass-scaled Jacobi vectors---as
opposed to the use of non-mass-scaled Jacobi vectors---simplifies 
the derivation of the fixed-$R$ matrix
elements for the ECG
basis functions (here, $R$ denotes the hyperradius;
see below and Sec.~\ref{sec_matrixelements}).
The Hamiltonian $H$ 
can now be written as a sum of the
relative Hamiltonian $H_{\rm{rel}}$ and the center
of mass Hamiltonian $H_{\rm{cm}}$,
\begin{eqnarray}
H = H_{\rm{rel}}+H_{\rm{cm}}
\end{eqnarray}
with
\begin{eqnarray}
H_{\rm{rel}}= 
T_{\rm{rel}}
+ V_{\rm{int}},
\end{eqnarray}
\begin{eqnarray}
T_{\rm{rel}} =
\sum_{j=1}^{n-1} -\frac{\hbar^2}{2 
} \nabla_{\vec{\rho}_j}^2
\end{eqnarray}
and 
\begin{eqnarray}
H_{\rm{cm}}= -\frac{\hbar^2}{2 
} \nabla_{\vec{\rho}_n}^2.
\end{eqnarray}
In the following, we seek solutions to the relative 
Schr\"odinger equation
\begin{eqnarray}
\label{eq_serel}
H_{\rm{rel}} \psi_E = E \psi_E,
\end{eqnarray} 
i.e., we seek to
determine
$\psi_E$ and $E$.
The energy $E$ can be negative or positive,
i.e., we consider both bound state and
scattering solutions.

We employ the hyperspherical coordinate 
approach~\cite{ritt11,linreview,macek,shertzer}, 
which has proven to
provide critical physical insights that, in some cases,
are more difficult or even impossible to unravel in alternative approaches.
The solution to the relative
Hamiltonian is divided into two steps: {\em{(i)}} the solution of a
Schr\"odinger like equation in the
hyperangular coordinates and {\em{(ii)}} the solution of
a  Schr\"odinger like equation in the
hyperradial coordinate. More specifically,
the idea is to expand the relative wave function
$\psi_E(\vec{\rho}_1,\cdots,\vec{\rho}_{n-1})$ in
terms of a complete set of hyperangular channel
functions $\Phi_{\nu}(R;\vec{\Omega})$ 
that depend parametrically on the hyperradius
$R$ 
and hyperradial weight functions
$F_{\nu E}(R)$~\cite{ritt11,linreview,macek,shertzer},
\begin{eqnarray}
\label{eq_wavehyperspherical}
\psi_E = R^{-(3n-4)/2} 
\sum_{\nu} F_{\nu E}(R) \Phi_{\nu}(R; \vec{\Omega}).
\end{eqnarray}
Here, $R$ denotes the hyperradius,
\begin{eqnarray}
R^2 = \sum_{k=1}^{n-1} 
\rho_k^2,
\end{eqnarray}
which 
has, as the
components of $\vec{\rho}_j$, units
of
``mass$^{1/2}$ times
length''. 
The mass-scaled hyperradius $R$ can be related to
the ``conventional unscaled hyperradius''
by pulling out a factor of $\sqrt{\mu}$, where $\mu$ is the hyperradial mass.
In Eq.~(\ref{eq_wavehyperspherical}),
$\vec{\Omega}$ collectively denotes the $3n-4$ hyperangles.
The hyperangles $\vec{\Omega}$ can be defined in different ways
(see Sec.~\ref{sec_matrixelements} for the definition employed in
this work). 

The channel functions
$\Phi_{\nu}(R;\vec{\Omega})$ form a complete set in the 
$(3n-4)$-dimensional Hilbert space
associated with the hyperangular degrees of 
freedom~\cite{ritt11,linreview,macek,shertzer}, 
\begin{eqnarray}
\int [\Phi_{\nu'}(R;\vec{\Omega})]^* 
\Phi_{\nu}(R;\vec{\Omega}) d^{3n-4} \vec{\Omega} = \delta_{\nu'\nu}.
\end{eqnarray}
The $\Phi_{\nu}(R;\vec{\Omega})$ are chosen to solve the
fixed-$R$ hyperangular Schr\"odinger equation
\begin{eqnarray}
\left[ H_{\rm{adia}} +V_{\rm{int}}(R,\vec{\Omega}) \right]
\Phi_{\nu}(R;\vec{\Omega}) = 
U_{\nu}(R) \Phi_{\nu}(R; \vec{\Omega}),
\end{eqnarray}
where
\begin{eqnarray}
H_{\rm{adia}}= T_{\Omega} +
V_{\rm{eff}}(R)
\end{eqnarray}
with
\begin{eqnarray}
T_{\Omega}= \frac{\hbar^2 \Lambda^2}{2  R^2} 
\end{eqnarray}
and
\begin{eqnarray}
V_{\rm{eff}}(R)=\frac{\hbar^2 (3n-4)(3n-6)}{8  R^2}.
\end{eqnarray}
The grandangular momentum operator $\Lambda$~\cite{ritt11,averybook} 
accounts for the kinetic energy
associated with the hyperangular degrees of freedom.
For our purposes, it proves advantageous
to define $T_{\Omega}$ through
\begin{eqnarray}
\label{eq_tomegasplit}
T_{\Omega}= T_{\rm{rel}} - T_R, 
\end{eqnarray}
where
\begin{eqnarray}
T_R = -\frac{ \hbar^2}{2} 
\frac{1}{R^{3n-4}} \frac{\partial}{\partial R} 
R^{3n-4} \frac{\partial}{\partial R} .
\end{eqnarray}
Inserting Eq.~(\ref{eq_wavehyperspherical})
into Eq.~(\ref{eq_serel}),
we find that the weight functions $F_{\nu E}(R)$ and
the relative energy $E$ are obtained by solving a set of
coupled hyperradial equations
\begin{eqnarray}
\left[ -\frac{\hbar^2}{2} \frac{\partial^2}{\partial R^2} 
+ U_{\nu}(R) 
\right]F_{\nu E}(R)
+V_{\rm{c},\nu}(R) 
= \nonumber \\
E F_{\nu E}(R),
\end{eqnarray}
where the coupling term
$V_{\rm{c},\nu}$ is given by
\begin{eqnarray}
V_{\rm{c},\nu}(R) = \nonumber \\
\sum_{\nu'} \left[
-2 P_{\nu \nu'}(R)  \frac{\partial F_{\nu' E}(R)}{\partial R}
- Q_{\nu \nu'}(R) F_{\nu' E}(R)
\right]
\end{eqnarray}
with
\begin{eqnarray}
P_{\nu \nu'}(R)= \frac{\hbar^2}{2}
\int [\Phi_{\nu}(R;\vec{\Omega})]^* 
\frac{\partial \Phi_{\nu'}(R;\vec{\Omega})}{\partial R}
d ^{3n-4} \vec{\Omega}
\end{eqnarray}
and
\begin{eqnarray}
Q_{\nu \nu'}(R)= \frac{\hbar^2}{2}
\int [\Phi_{\nu}(R;\vec{\Omega})]^* 
\frac{\partial^2 \Phi_{\nu'}(R;\vec{\Omega})}{\partial R^2}
d ^{3n-4} \vec{\Omega}.
\end{eqnarray}

To reiterate,
the hyperspherical framework consists of two steps:
In the first step, the $(3n-4)$-dimensional hyperangular Schr\"odinger
equation is solved, yielding
$U_{\nu}(R)$, $P_{\nu \nu'}(R)$ and $Q_{\nu \nu'}(R)$.
In the second step, the coupled set of one-dimensional
hyperradial equations
is solved, yielding $F_{\nu E}(R)$ and $E$.
This paper focuses primarily
on solving the hyperangular Schr\"odinger
equation.
Expressions for the relevant matrix elements,
valid for any $n$, are derived
and applications to systems with $n=4$ 
are presented.

\section{Functional form of the basis functions}
\label{sec_basisfunctions}
To solve the hyperangular Schr\"odinger equation,
we expand the channel functions
$\Phi_{\nu}(R; \vec{\Omega})$
for fixed $R$
in terms of ECG
basis functions 
$\psi_k(\underline{A}^{(k)},\vec{u}_{1}^{(k)},\vec{u}_{2}^{(k)},\vec{x})|_R$,
\begin{eqnarray}
\label{eq_basissetexpansion}
\Phi_{\nu}(R;\vec{\Omega}) = 
\sum_{k=1}^{N_b} 
c _k {\cal{S}} 
\psi_k(\underline{A}^{(k)},\vec{u}_{1}^{(k)},\vec{u}_{2}^{(k)},\vec{x})|_R,
\end{eqnarray}
where $\vec{x}$ collectively denotes
the $3n-3$ Jacobi vectors,
$\vec{x}=(\vec{\rho}_1,\cdots,\vec{\rho}_{n-1})$.
In Eq.~(\ref{eq_basissetexpansion}), the notation ``$|_R$'' indicates
that $\psi_k$ is evaluated at a fixed hyperradius $R$.
The $(n-1) \times (n-1)$ dimensional matrices
$\underline{A}^{(k)}$ 
are symmetric and positive definite. The $n(n-1)/2$ independent
elements of $\underline{A}^{(k)}$ are treated as variational parameters
of the $k$$^{\rm{th}}$ basis function.
The elements of the $(n-1)$-dimensional
vectors
$\vec{u}_{1}^{(k)}$ and $\vec{u}_{2}^{(k)}$
are also treated as variational parameters.
The optimization scheme employed to determine the values of
these variational
parameters is discussed in Sec.~\ref{sec_implementation}.
In Eq.~(\ref{eq_basissetexpansion}), ${\cal{S}}$ denotes
an operator that imposes the proper symmetry under exchange
of identical particles.
For the three-body 
system consisting of two identical fermions and an impurity,
e.g.,
we have ${\cal{S}}=1-P_{12}$.
For the four-body 
system consisting of three identical fermions and an impurity
(see Sec.~\ref{sec_applications}),
we have ${\cal{S}}=1-P_{12}-P_{13}-P_{23}+P_{12}P_{23}+P_{13}P_{32}$.

The functional form of the basis functions depends on the 
$L^{\Pi}$ symmetry considered.
We consider ECG basis functions 
of the form~\cite{cgbook,varg95,varg98a,suzu98,suzu00,suzu08}
\begin{eqnarray}
\label{eq_basisfunction}
\psi(\underline{A},\vec{u}_1,\vec{u}_2,\vec{x})=
N_{L \Pi} \times \nonumber \\
|\vec{v}_1|^{l_1} |\vec{v}_2|^{l_2}
[Y_{l_1}(\hat{v}_1) \otimes Y_{l_2} (\hat{v}_2)]_{LM_L} 
\exp \left( 
-\frac{\vec{x}^T \underline{A} \vec{x}}{2}
\right),
\end{eqnarray}
which can be used to describe states with
$0^+,1^+,1^-,2^+,\cdots$ symmetry but not states with $0^-$ 
symmetry~\cite{footnotezerominus}.
The three-dimensional vectors $\vec{v}_j$, 
$j=1$ and 2,
are defined through
\begin{eqnarray}
\label{eq_vvector}
\vec{v}_{j}=\left( \vec{u}_j  \right)^T \vec{x}=
\sum_{k=1}^{n-1} u_{j,k} \vec{\rho}_k,
\end{eqnarray}
where $u_{j,k}$ denotes the $k$$^{th}$ component of the
vector $\vec{u}_j$.
We denote the elements of the vector $\vec{v}_j$ by
$v_{j,1}$, $v_{j,2}$ and $v_{j,3}$.
In Eq.~(\ref{eq_basisfunction}), 
the notation $[Y_{l_1} \otimes Y_{l_2 }]_{LM_L}$
indicates that the two spherical harmonics $Y_{l_j m_j}$
are coupled to a function with total angular momentum $L$
and projection quantum number $M_L$,
and $N_{L \Pi}$ denotes a normalization constant.
 
In the following, we write the basis functions out for
$M_L=0$; in writing the basis functions for a specific 
symmetry, we choose the normalization constant $N_{L \Pi}$
in Eq.~(\ref{eq_basisfunction}) conveniently. 
Throughout this paper,
we restrict ourselves to states with $0^+,1^-$ and $1^+$ symmetry.
The $L^{\Pi}=0^+$ basis functions, obtained by setting
$l_1$ and $l_2$ to $0$, are independent of
$\vec{u}_1$ and $\vec{u}_2$ 
(or, equivalently, of $\vec{v}_1$ and $\vec{v}_2$),
\begin{eqnarray}
\label{eq_basis_zeroplus}
\psi(\underline{A},\vec{x}) =
\exp \left( -\frac{\vec{x}^T \underline{A} \vec{x} }{2} \right).
\end{eqnarray}
The $L^{\Pi}=1^-$ basis 
functions, obtained by setting $l_1$ to $1$ and $l_2$ to $0$, 
depend on $\vec{u}_1$ but not on $\vec{u}_2$,
\begin{eqnarray}
\label{eq_basis_oneminus}
\psi(\underline{A}, \vec{u}_{1},\vec{x}) =
3^{1/2} v_{1,3}
\exp \left( -\frac{\vec{x}^T \underline{A} \vec{x} }{2} \right).
\end{eqnarray}
Lastly, the $L^{\Pi}=1^+$ basis functions, obtained by setting
$l_1$ and $l_2$ to $1$, depend on both $\vec{u}_1$
and $\vec{u}_2$,
\begin{eqnarray}
\label{eq_basis_oneplus}
\psi(\underline{A},\vec{u}_1,\vec{u}_2,\vec{x}) = \nonumber \\
\frac{3}{2^{1/2}} 
\left( v_{1,2} v_{2,1} - v_{1,1} v_{2,2} \right)
\exp \left( -\frac{\vec{x}^T \underline{A} \vec{x} }{2} \right).
\end{eqnarray}

The next section presents relatively compact expressions for the 
fixed-$R$ matrix elements 
between two
basis functions 
$\psi=\psi(\underline{A},\vec{u}_{1},\vec{u}_{2},\vec{x})$
and
$\psi'=\psi(\underline{A}',\vec{u}_{1}',\vec{u}_{2}',\vec{x})$.
Throughout, we assume that $\psi$ and $\psi'$
are characterized by the same $L$, $M_L$ and $\Pi$ quantum numbers.
For systems with finite angular momentum, 
neither the fixed-$R$ overlap matrix element nor the 
fixed-$R$ matrix elements
for $T_{\Omega}$, $P_{\nu \nu'}$, 
$Q_{\nu \nu'}$, and  $V_{\rm{int}}$
have, to the best of our knowledge,
been reported in the literature.

\section{Matrix elements for hyperspherical explicitly
correlated Gaussians}
\label{sec_matrixelements}
We introduce the short-hand notation
\begin{eqnarray}
\label{eq_overlapdef}
\langle \psi' |\psi \rangle|_R = \nonumber \\
\int [\psi( \underline{A}' ,\vec{u}_1',\vec{u}_2',\vec{x}) |_R ]^*
\psi( \underline{A},\vec{u}_1,\vec{u}_2,\vec{x})|_R d^{3n-4} \vec{\Omega},
\end{eqnarray}
\begin{eqnarray}
\label{eq_pcouplingdef}
\langle \psi' |P|\psi \rangle|_R = \nonumber \\
\frac{\hbar^2}{2}
\int [\psi( \underline{A}' ,\vec{u}_1',\vec{u}_2',\vec{x}) |_R ]^*
\frac{\partial \psi( \underline{A},\vec{u}_1,\vec{u}_2,\vec{x})}
{\partial R} \Big|_R d^{3n-4} \vec{\Omega},
\end{eqnarray}
\begin{eqnarray}
\label{eq_qcouplingdef}
\langle \psi' |Q|\psi \rangle|_R = \nonumber \\
\frac{\hbar^2}{2}
\int [\psi( \underline{A}' ,\vec{u}_1',\vec{u}_2',\vec{x}) |_R ]^*
\frac{\partial^2 \psi( \underline{A},\vec{u}_1,\vec{u}_2,\vec{x})}
{\partial R^2} \Big|_R d^{3n-4} \vec{\Omega},
\end{eqnarray}
and
\begin{eqnarray}
\label{eq_kinetichyperdef}
\langle \psi' | T_{\Omega} | \psi \rangle |_R = \nonumber \\
\frac{1}{2} 
\int [\psi( \underline{A}' ,\vec{u}_1',\vec{u}_2',\vec{x}) |_R ]^*
T_{\Omega}
\psi( \underline{A},\vec{u}_1,\vec{u}_2,\vec{x}) |_R d^{3n-4} \vec{\Omega}+
\nonumber \\
\frac{1}{2}
\int [\psi( \underline{A} ,\vec{u}_1,\vec{u}_2,\vec{x}) |_R ]^*
T_{\Omega}
\psi( \underline{A}',\vec{u}_1',\vec{u}_2',\vec{x}) |_R d^{3n-4} \vec{\Omega}
.
\end{eqnarray}
As in Refs.~\cite{vonstecherPRA,vonstecherThesis,ritt11},
Eq.~(\ref{eq_kinetichyperdef}) explicitly symmetrizes the
matrix element associated with the hyperangular kinetic energy.

To evaluate the fixed-$R$
matrix elements defined in 
Eqs.~(\ref{eq_overlapdef})-(\ref{eq_kinetichyperdef}),
we introduce
a new set of coordinates
$\vec{y}$,
$\vec{y}=(\vec{y}_1,\cdots,\vec{y}_{n-1})$,
through
$\vec{y}=(\underline{U}_B)^T \vec{x}$.
The matrix $\underline{U}_B$ is chosen such that
the matrix $(\underline{U}_B)^T \underline{B} \underline{U}_B$,
where 
\begin{eqnarray}
\label{eq_bmatrix}
\underline{B} = \underline{A} + \underline{A}',
\end{eqnarray}
is
diagonal with diagonal elements $\beta_1,\cdots,\beta_{n-1}$, i.e.,
such that
$\vec{x}^T \underline{B} \vec{x}= \sum_{j=1}^{n-1} \beta_j \vec{y}_j^2$.
It follows that the arguments of the exponentials in 
the integrals defined in Eqs.~(\ref{eq_overlapdef})-(\ref{eq_kinetichyperdef})
reduce to quadratic forms.
Since the coordinate transformation from $\vec{x}$ to $\vec{y}$ 
is orthogonal,
we have {\em{(i)}} 
$R^2 = \sum_{j=1}^{n-1} \vec{y}_j^2$ and
{\em{(ii)}} 
$\int \cdots d^{3(n-1)} \vec{x} = \int \cdots d^{3(n-1)} \vec{y}$.

To perform the integration over $\vec{\Omega}$,
we need to specify the $3n-4$ hyperangles.
Following 
Refs.~\cite{russian,bohn98,vonstecherPRA,vonstecherThesis,ritt11}, 
we define $2(n-1)$ angles
as the polar and azimuzal angles $\vartheta_j$ and
$\varphi_j$ of the $n-1$ vectors $\vec{y}_j$.
The remaining $n-2$ hyperangles $\gamma_1,\cdots,\gamma_{n-2}$ are defined
in terms of the direction of the $(n-1)$-dimensional vector
$\vec{s}$, 
where $\vec{s}=(y_1,\cdots,y_{n-1})$ and $y_j=|\vec{y}_j|$.
Specifically, 
we write
\begin{eqnarray}
\label{eq_definition_hyperangles}
y_1&=&R \sin \gamma_1 \sin \gamma_2 \cdot \cdots \cdot \sin \gamma_{n-2},
\nonumber \\
y_2&=&R \cos \gamma_1 \sin \gamma_2 \cdot \cdots \cdot \sin \gamma_{n-2},
\nonumber \\
y_3&=&R \cos \gamma_2 \sin \gamma_3 \cdot \cdots \cdot \sin \gamma_{n-2},
\nonumber \\
&&\cdots,
\nonumber \\
y_{n-2}&=&R \cos \gamma_{n-3} \sin \gamma_{n-2},
\nonumber \\
y_{n-1} &=& R \cos \gamma_{n-2},
\end{eqnarray}
where $\gamma_j \in [0, \pi/2]$.
This restriction on the range of the angles ensures that the
$y_j$ are non-negative.
Correspondingly,
we have~\cite{russian,bohn98}
\begin{eqnarray}
\label{eq_hyperangularvolumeelement}
\int \cdots d^{3n-4} \vec{\Omega} = \nonumber \\
\int \cdots
\left( \prod_{j=1}^{n-1} d^2 \hat{y}_j \right)
\left( \prod_{k=1}^{n-2} \sin^{3k-1} \gamma_k \cos ^2 \gamma_k d \gamma_k
\right)
,
\end{eqnarray}
where $d^2 \hat{y}_j$ denotes the ``usual'' angular piece 
of the volume element in spherical coordinates,
$d^2 \hat{y}_j= \sin \vartheta_j d \vartheta_j d \varphi_j$.

In general, the expressions for the matrix
elements defined in Eqs.~(\ref{eq_overlapdef})-(\ref{eq_kinetichyperdef})
depend on the functional form of the basis functions 
$\psi(\underline{A},\vec{u}_1,\vec{u}_2,\vec{x})$
(see Sec.~\ref{sec_basisfunctions}).
For the basis functions defined in 
Eqs.~(\ref{eq_basis_zeroplus})-(\ref{eq_basis_oneplus}),
the integrations involving the angles
$\vartheta_j$ and $\varphi_j$ ($j=1,\cdots,n-1$) can be performed 
analytically, yielding
\begin{eqnarray}
\label{eq_overlap_firstint}
\langle \psi' |\psi \rangle |_R =
(4 \pi)^{n-1}
\int 
f_o(\vec{s})
\exp\left(- \frac{1}{2} \sum_{j=1}^{n-1} \beta_j y_j^2 \right) \nonumber \\
\left( \prod_{k=1}^{n-2} \sin^{3k-1} \gamma_k \cos ^2 \gamma_k d \gamma_k
\right),
\end{eqnarray}
\begin{eqnarray}
\label{eq_p_firstint}
\langle \psi' | P | \psi \rangle |_R =
-\frac{\hbar^2(4 \pi)^{n-1}}{2} \int 
f_{P}(\vec{s})
\exp\left(- \frac{1}{2} \sum_{j=1}^{n-1} \beta_j y_j^2 \right) 
\nonumber \\
\left( \prod_{k=1}^{n-2} \sin^{3k-1} \gamma_k \cos ^2 \gamma_k d \gamma_k
\right),
\end{eqnarray}
\begin{eqnarray}
\label{eq_q_firstint}
\langle \psi' | Q | \psi \rangle |_R =
-\frac{\hbar^2(4 \pi)^{n-1}}{2} \int 
f_{Q}(\vec{s})
\exp\left(- \frac{1}{2} \sum_{j=1}^{n-1} \beta_j y_j^2 \right) 
\nonumber \\
\left( \prod_{k=1}^{n-2} \sin^{3k-1} \gamma_k \cos ^2 \gamma_k d \gamma_k
\right),
\end{eqnarray}
and
\begin{eqnarray}
\label{eq_kinetic_firstint}
\langle \psi' | T_{\Omega} | \psi \rangle |_R =
-\frac{\hbar^2(4 \pi)^{n-1}}{4} \int 
f_{\Omega}(\vec{s})
\exp\left(- \frac{1}{2} \sum_{j=1}^{n-1} \beta_j y_j^2 \right) 
\nonumber \\
\left( \prod_{k=1}^{n-2} \sin^{3k-1} \gamma_k \cos ^2 \gamma_k d \gamma_k
\right).
\end{eqnarray}
The matrix elements $\langle \psi'| \psi \rangle |_R$,
$\langle \psi'|P| \psi \rangle |_R$, $\langle \psi' |Q| \psi \rangle |_R$
and $\langle \psi'| T_{\Omega}|\psi \rangle |_R$, 
Eqs.~(\ref{eq_overlap_firstint})-(\ref{eq_kinetic_firstint}), 
have been written such that 
$f_o(\vec{s})$,
$f_P(\vec{s})$,
$f_Q(\vec{s})$
and $f_{\Omega}(\vec{s})$ have
analogous functional
forms.
We write
\begin{eqnarray}
\label{eq_flambdageneral}
f_{o}(\vec{s})&=&
d^{(0)}+\sum_{j=1}^{n-1} 
\left[d^{(2)}_j y_{j}^2+d^{(4)}_j y_{j}^4+d^{(6)}_j y_{j}^6
\right]+
\nonumber
\\
&&\sum _{k>j=1}^{n-1} \left[
d^{(22)}_{j,k} y_{j}^2y_{k}^2+ d^{(44)}_{j,k} y_j^4 y_k^4 \right] +
\nonumber
\\
&&\sum _{j=1,k=1,k\neq j}^{n-1} 
\left[ d^{(24)}_{j,k} y_{j}^2 y_{k}^4 +
d^{(26)}_{j,k} y_j^2 y_k^6 \right] +
\nonumber \\
&&\sum _{k>j=1;l \ne j,k}^{n-1} \left[
d^{(222)}_{j,k,l}y_{j}^2y_{k}^2y_{l}^2 +
d^{(224)}_{j, k,l} y_j^2 y_k^2 y_l^4 \right]+
\nonumber \\
&&\sum_{k>j=1;l>j;m>l;m \ne l\ne k \ne j} ^{n-1}
d_{j,k,l,m}^{(2222)} y_j^2 y_k^2 y_l^2 y_m^2
.
\end{eqnarray}
In writing $f_{o}(\vec{s})$, we dropped terms that contain odd powers
of $y_j$ since these terms average to zero when 
integrating over the remaining hyperangles.
The quantities 
$f_P(\vec{s})$, $f_Q(\vec{s})$ and $f_{\Omega}(\vec{s})$
are obtained by replacing the $d$'s in Eq.~(\ref{eq_flambdageneral}) by
$p$'s, $q$'s and $b$'s, respectively.
The super- and subscripts of the $d$-, $p$-, $q$- 
and $b$-coefficients indicate 
the polynomial in the $y$'s that the coefficients are associated with.
The $d$-, $p$-, $q$- and $b$-coefficients 
depend on the symmetry of the wave function, and are listed 
in Appendix~\ref{appendix_symmetry}
for states with $L^{\Pi}=0^+,1^-$ and $1^+$ symmetry.
It should be noted that, depending 
on the symmetry and number of particles, a varying number of the
$d$-, $p$-, $q$- and $b$-coefficients vanish.
Appendix~\ref{appendix_derivation}
exemplarily
illustrates how to obtain Eq.~(\ref{eq_overlap_firstint}) 
for the three-body system with $L^{\Pi}=1^-$ 
symmetry.
We emphasize that 
Eqs.~(\ref{eq_overlap_firstint})-(\ref{eq_flambdageneral}),
together with the expressions given in Appendix~\ref{appendix_symmetry},
apply to any number of particles.
For states with $L>1$ and $L^{\Pi}=0^-$,
the definitions of the $d$-, $p$-, $q$- and $b$-coefficients 
contained in $f_o$,
$f_P$, $f_Q$ and $f_{\Omega}$ change and polynomials
in the $y$'s of higher power than those considered
in Eq.~(\ref{eq_flambdageneral}) may appear.

The integration over $\gamma_1$ 
in Eqs.~(\ref{eq_overlap_firstint})-(\ref{eq_kinetic_firstint})
can also be done analytically.
To perform this integration, we recognize that the hyperangle
$\gamma_1$ enters into $y_1$ and $y_2$ but not into
$y_j$ with $j \ge 3$. 
Using Eq.~(\ref{eq_definition_hyperangles}),
we write $y_1^2= R^2 \sin ^2 \gamma_1 H(\gamma_2,\cdots,\gamma_{n-2})$
and
$y_2^2= R^2 \cos ^2 \gamma_1 H(\gamma_2,\cdots,\gamma_{n-2})$,
where 
$H(\gamma_2,\cdots,\gamma_{n-2})=1$  for $n=3$ and
\begin{eqnarray}
H(\gamma_2,\cdots,\gamma_{n-2})=
\sin^2 \gamma_2 \times \cdots \times \sin^2 \gamma_{n-2}
\end{eqnarray}
for $n>3$.
In the following, we suppress the dependence of $H$
on the angles $\gamma_j$ ($j \ge 2$)
and rewrite 
$f_{o}(\vec{s})$ 
such that the dependence on $\gamma_1$ is ``isolated'',
\begin{eqnarray}
\label{eq_sc_coeff}
f_{o}(\vec{s})&=&
sc_{00} +
sc_{20} H R^2 \sin^2 \gamma_1 + sc_{02} H R^2 \cos^2 \gamma_1 +
\nonumber \\
&&sc_{40} (H R^2)^2 \sin^4 \gamma_1 + sc_{04} (H R^2)^2 \cos^4 \gamma_1 +
\nonumber \\
&&sc_{60} (H R^2)^3 \sin^6 \gamma_1 + sc_{06} (H R^2)^3 \cos^6 \gamma_1 + 
\nonumber \\
&&sc_{22} (H R^2)^2 \sin^2 \gamma_1 \cos^2 \gamma_1 + 
\nonumber \\
&&sc_{44} (H R^2)^4 \sin^4 \gamma_1 \cos^4 \gamma_1 + 
\nonumber \\
&&sc_{24} (H R^2)^3 \sin^2 \gamma_1 \cos^4 \gamma_1 + 
\nonumber \\
&&sc_{42} (H R^2)^3 \sin^4 \gamma_1 \cos^2 \gamma_1 + 
\nonumber \\
&&sc_{26} (H R^2)^4 \sin^2 \gamma_1 \cos^6 \gamma_1 + 
\nonumber \\
&&sc_{62} (H R^2)^4 \sin^6 \gamma_1 \cos^2 \gamma_1.
\end{eqnarray}
The coefficients $sc_{jk}$
depend on the hyperangles $\gamma_l$ with $l \ge 2$ and the
$d$-coefficients, and
are defined in Appendix~\ref{appendix_general}.
The subscripts $j$ and $k$ of the $sc$-coefficients
denote respectively the powers of $\sin \gamma_1$ and $\cos \gamma_1$ 
that the coefficients $sc_{jk}$ are associated with.
Using Eq.~(\ref{eq_sc_coeff}), we find
\begin{eqnarray}
\label{eq_integralovergammaone}
\int_0^{\pi/2} f_{o}(\vec{s}) \exp \left(
-\frac{1}{2} \sum_{j=1}^{n-1} \beta_j y_j^2 \right) \sin^2 \gamma_1
\cos^2 \gamma_1 d \gamma_1
= \nonumber \\
\frac{\pi}{16 \zeta} 
\exp \left( -\frac{1}{4} H R^2 (\beta_1+ \beta_2) - \frac{1}{2} 
\sum_{j=3}^{n-1} \beta_j y_j^2
\right) \times \nonumber 
\\
\left[
M_1 \frac{I_1(\zeta)}{\zeta}
+ M_2 \frac{I_2(\zeta)}{\zeta}
\right],
\end{eqnarray}
where $I_1$ and $I_2$ denote Bessel functions
and $\zeta$ is defined through
\begin{eqnarray}
\label{eq_xi}
\zeta = \frac{1}{4} (\beta_1 - \beta_2) H R^2.
\end{eqnarray}
The quantities $M_1$ and $M_2$ can be written in terms of
$RH^2$, $\zeta$ and the $sc$-coefficients;
$M_1$ and $M_2$ depend on
$\gamma_2,\cdots,\gamma_{n-2}$
since $H$ and the $sc$-coefficients depend on these angles.
Explicit expressions for $M_1$ and $M_2$ are given in
Appendix~\ref{appendix_general}.
Using Eq.~(\ref{eq_integralovergammaone}) in 
Eq.~(\ref{eq_overlap_firstint}),
the matrix element
$\langle \psi | \psi \rangle|_R$ is known fully analytically for $n=3$
and reduces to a $(n-3)$-dimensional integral for $n>3$.
For $n=4$ and 5, the remaining one- and two-dimensional
integrations can, as discussed in Sec.~\ref{sec_implementation},
be
performed numerically with high accuracy.
Expressions~(\ref{eq_sc_coeff}) and (\ref{eq_integralovergammaone})
also apply to
$f_{P}(\vec{s})$, $f_{Q}(\vec{s})$ and $f_{\Omega}(\vec{s})$
if the $d$-coefficients are replaced by the $p$-, $q$- and $b$-coefficients,
respectively.

For $L^{\Pi}=0^+$,
our results obtained using the
above expressions agree with those reported in 
Refs.~\cite{vonstecherPRA,vonstecherThesis} for $n=3$ and 4.
Motivated by our desire to express the various matrix elements for
different 
number of particles
$n$ and $L^{\Pi}$ symmetries
in a unified framework, we adopted a notation
that differs notably
from the notation adopted in 
Refs.~\cite{vonstecherPRA,vonstecherThesis,ritt11,vonstecherCorrection}.

We now turn to the evaluation of the interaction matrix element.
We define
\begin{eqnarray}
\label{eq_twobodydef}
\langle \psi' | V_{kl}(\vec{r}_{kl})| 
\psi \rangle |_R  = \nonumber \\
\int [\psi( \underline{A}' ,\vec{u}_1',\vec{u}_2',\vec{x}) |_R ]^*
V_{kl}(\vec{r}_{kl})
\psi( \underline{A},\vec{u}_1,\vec{u}_2,\vec{x}) |_R d^{3n-4} \vec{\Omega}.
\end{eqnarray}
If the two-body potential
$V_{kl}(\vec{r}_{kl})$ is parameterized
by a spherically symmetric 
short-range Gaussian $V_{\rm{g}}(r_{kl})$ with depth $D_{kl}$
and range $r_{0,kl}$,
\begin{eqnarray}
V_{\rm{g}}(r_{kl}) = -D_{kl} \exp \left[ 
-\left( \frac{r_{kl}}{\sqrt{2} r_{0,kl}}
\right)^2 \right],
\end{eqnarray}
then
$\langle \psi'| V_{\rm{g}}(r_{kl}) | \psi \rangle |_R$ 
is equivalent to $- D_{kl} \langle \psi'| \psi \rangle |_R$
if $\underline{A}$ and $\underline{A}'$ are replaced by
$\underline{A}+\underline{W}^{(kl)}/(2 r_{0,kl}^2)$ and
$\underline{A}'+\underline{W}^{(kl)}/(2 r_{0,kl}^2)$,
respectively.
Here,
the matrix $\underline{W}^{(kl)}$
is defined as
\begin{eqnarray}
\underline{W}^{(kl)} = \vec{\omega}^{(kl)} \left(
\vec{\omega}^{(kl)} \right)^T,
\end{eqnarray}
where the vector $\vec{\omega}^{(kl)}$
provides the transformation
from the Jacobi coordinates $\vec{x}$ to the interparticle distance
vector $\vec{r}_{kl}$,
\begin{eqnarray}
\vec{r}_{kl} = \left(  \vec{\omega} ^{(kl)} \right) ^T \vec{x}.
\end{eqnarray}
It follows that we can use 
Eq.~(\ref{eq_overlap_firstint}) [see also
Eqs.~(\ref{eq_flambdageneral})-(\ref{eq_xi})]
if $\zeta$ 
is replaced by $\zeta^{(kl)}$
and if the $\beta_j$ are replaced by
$\beta_j^{(kl)}$.
Similar to $\zeta$,
$\zeta^{(kl)}$ is defined through
\begin{eqnarray}
\zeta^{(kl)} = 
\frac{1}{4} (\beta_1^{(kl)} - \beta_2^{(kl)}) H R^2
\end{eqnarray}
and
the
$\beta_j^{(kl)}$ 
denote the eigenvalues of the matrix
$\underline{B}+\underline{W}^{(kl)}/r_{0,kl}^2$.

\section{Implementation details of the HECG approach
and proof-of-principle applications}
\label{sec_implementation}
As discussed in the previous sections, the determination of the
effective hyperradial potential curves and coupling matrix elements
requires that the hyperangular Schr\"odinger
equation be solved for several hyperradii $R$. For each fixed $R$,
the determination of the linear and non-linear variational
parameters is accomplished following approaches very similar to those 
employed in the ``standard'' (non fixed $R$) ECG 
approach~\cite{cgbook}.
For a given set of basis functions, and thus 
for a given set of non-linear
variational
parameters, the expansion coefficients $c_k$, $k=1,\cdots, N_b$
[see Eq.~(\ref{eq_basissetexpansion})],
are determined by diagonalizing the generalized
eigenvalue problem defined by the fixed-$R$ Hamiltonian matrix
and the fixed-$R$ overlap matrix. According to the 
generalized Ritz variational principle, the 
$N_b$ eigenvalues provide variational upper bounds to
the exact eigenvalues of the hyperangular
Schr\"odinger equation. 

The non-linear 
variational parameters are collected in $\underline{A}^{(k)}$,
$\vec{u}_1^{(k)}$ and $\vec{u}_2^{(k)}$, where $k=1,\cdots,N_b$,
and
determined using the stochastic variational
approach~\cite{svm}, i.e., through a trial and error procedure. 
For concreteness, we consider the situation where we aim
to determine the energetically lowest
lying hyperangular eigenvalue $U_0(R)$ for a given
$R$ value.
We start with a small basis set (typically consisting of
just one basis function). To add the next basis function,
we semi-randomly generate $N_{\rm{T}}$ trial basis 
functions ($N_{\rm{T}}$ is typically of the order of a few thousand), 
yielding 
$N_{\rm{T}}$ trial basis sets. We determine the eigenvalue for each of these
trial basis sets and 
choose the trial basis set that yields the smallest eigenvalue 
as the new basis set. 
Following the same selection process,
we continue to enlarge the basis set one
basis function at a time.
This procedure is repeated till the basis set is sufficiently
complete and the desired accuracy
of the energetically lowest lying eigenvalue
is reached. The optimization
of excited states proceeds analogously. If nearly degenerate states exist, it
is advantageous to simultaneously minimize the eigenvalues of multiple
states. 

Since the trial and error procedure 
``throws out'' most of the trial basis functions generated, the
resulting basis set is typically comparatively small. 
For the four-body systems discussed below,
e.g., we achieve convergence for $N_b$ of the order of 100. 
Moreover, we have found that a carefully constructed basis set avoids 
a number of problems that can arise from the fact that the basis functions
are not orthogonal.
In particular, the minimization scheme that underlies
the trial and error procedure
tends to select basis functions
that have relatively small overlaps among each other, i.e., that are 
``fairly 
linearly independent''. In some cases, however, 
the trial and error procedure does not
fully eliminate numerical issues arising from the linear dependence of
the basis functions.
Thus, we add another check and
reject a given  trial basis function $\psi_{\rm{T}}$
if its overlap with one or more of the basis functions already selected
is too large,
i.e., if the quantity
$\langle \psi_{\rm{T}} | \psi_k \rangle|_R$
is larger than a preset value $\epsilon$, where we 
normalize $\psi_{\rm{T}}$ and $\psi_k$ such that
$\langle \psi_{\rm{T}} | \psi_{\rm{T}} \rangle|_R=
\langle \psi_k | \psi_k \rangle|_R=1$.
We have used $\epsilon=0.9$ or $0.95$ in most of
the calculations reported below.
A similar criterion is employed in the 
HECG approach of Ref.~\cite{vonstecherPRA}
and in the standard ECG approach~\cite{cgbook}. 

We now discuss the determination
of the non-linear parameters that characterize
the basis functions. The selection of the parameters
of the matrices $\underline{A}^{(k)}$ is guided by physical 
considerations.
The fact that
the basis functions have 
to be ``compatible'' with the value of the hyperradius
considered implies that the parameters of
the matrices $\underline{A}^{(k)}$ have to be chosen such that the
quantity $\sum_{j<l}^n(d_{jl}^{(k)})^2$ is of the order of $R^2/\mu$. 
The
width parameters $d_{jl}^{(k)}$ are related to
the parameter matrix $\underline{A}^{(k)}$ via 
\begin{eqnarray}
\vec{x}^T \underline{A}^{(k)} \vec{x} = 
\sum_{j<l} \frac{r_{jl}^2}{(d_{jl}^{(k)})^2}.
\end{eqnarray}
Moreover, the basis functions have to 
govern the dynamics that occurs at the length scale of the 
two-body interaction
potential. Correspondingly, 
we consider different types of basis functions:
The first type is characterized by all $d_{jl}^{(k)}$ being of the 
order of $R/(n\sqrt{\mu})$;
the second type is characterized by one $d_{jl}^{(k)}$ being of
the order of the range of the underlying two-body potential
(we use $r_0$ to denote the smallest of the $r_{0,jl}$'s) 
and all other $d_{jl}^{(k)}$ being of the order of 
$R/(n \sqrt{\mu})$;
the third type is characterized by two $d_{jl}^{(k)}$ being of
the order of $r_0$ and all other $d_{jl}^{(k)}$ being of the order of 
$R/(n \sqrt{\mu})$;
and so on.
The exact values of the $d_{jl}^{(k)}$ are chosen stochastically
from pre-defined parameter windows, which are chosen
according to the above considerations.
We have found that the convergence of the hyperangular eigenvalues
for strongly interacting systems with large $R/(\sqrt{\mu}r_0)$
depends quite sensitively on the choice of the 
parameter windows.
As in Ref.~\cite{vonstecherPRA},
we allow for basis functions with positive and negative widths. 
The elements of the parameter vectors $\vec{u}_{1}^{(k)}$
and $\vec{u}_{2}^{(k)}$ are
chosen to lie between $-1$ and $1$.

The quantity $|\xi|$, see Eq.~(\ref{eq_xi}), takes on 
large values if $|\beta_1| \gg |\beta_2|$
or
$|\beta_2| \gg |\beta_1|$.
This situation arises quite frequently if the hyperradius $R$ is
much larger than $\sqrt{\mu}r_0$ 
and
causes numerical difficulties when
evaluating the Bessel functions.
We mitigate these difficulties 
as follows.
For concreteness, we consider the $n=4$ case and the
integral involving $I_1(\xi)$.
For large $|\xi|$,
we write the integral over the hyperangle $\gamma_2$
[see Eqs.~(\ref{eq_overlap_firstint}) 
and (\ref{eq_integralovergammaone})] as
\begin{eqnarray}
\label{eq_numericalint}
\int_0^1 \xi^{-1}
\exp 
\left[ -\frac{1}{4}(\beta_1+\beta_2) R^2(1-x^2)
-\frac{1}{2}\beta_3 R^2 x^2
\right] \times \nonumber \\
M_1(x) \frac{I_1(\xi)}{\xi}
(1-x^2)^2 x^2 dx 
\approx \nonumber \\
(2 \pi)^{-1/2}
\times \nonumber \\
\int_0^1 
\xi^{-1} \exp
\left[
- \frac{1}{2} 
\min(\beta_1,\beta_2) R^2 (1-x^2) 
-\frac{1}{2}  \beta_3 R^2 x^2
\right] \nonumber \\
\left( 
\frac{1}{|\xi|^{3/2}} - \frac{3}{8 |\xi|^{5/2}}
- \frac{15}{128 |\xi|^{7/2}}
- \frac{105}{1024 |\xi|^{9/2}}
\right) \times \nonumber \\
M_1(x)
(1-x^2)^2 x^2 dx ,
\end{eqnarray}
where $x=\cos \gamma_2$ [implying $H(\gamma_2)= 1-x^2$].
We use the right hand side of Eq.~(\ref{eq_numericalint}) when $|\xi|>50$.
An analogous expansion is done for the part of the integrand
that involves $I_2(\xi)$.

Although $M_1$ depends on $x$, the overall behavior of the 
integrand in Eq.~(\ref{eq_numericalint}) is determined by the exponential.
In particular, depending on the signs of 
$\min(\beta_1,\beta_2)$ and $\beta_3$, the integrand 
can be sharply peaked 
at $x \approx 0$ or $x \approx 1$.
To perform the integration over $x$ (i.e., the integration
over the hyperangle $\gamma_2$)
numerically, 
we divide the 
integral into 
three sectors: the left, middle and right sectors.
The sector boundaries $x_{\rm{mid},1}$ and $x_{\rm{mid},2}$
are determined dynamically depending on the behavior
of the integrand.
The default values are $x_{\rm{mid},1}=0.3$ and $x_{\rm{mid},2}=0.7$.
However, if $\beta_3 R^2 > \delta$
or $\min(\beta_1,\beta_2) R^2 <-\delta$, we choose 
$x_{\rm{mid},1}=1/\sqrt{ \max(\beta_3 R^2,|\min(\beta_1,\beta_2)|R^2)}$.
Similarly, if $\beta_3 R^2 < -\delta$
or $\min(\beta_1,\beta_2) R^2 >\delta$, we choose 
$x_{\rm{mid},2}=1-1/\sqrt{ \max(|\beta_3| R^2,\min(\beta_1,\beta_2)R^2)}$.
Our calculations reported below use 
$\delta=20$.
The numerical integration for each sector is performed using
a standard Gauss-Laguerre integration scheme~\cite{numericalrecipe}.
Typically, we choose the same order $N_{\rm{order}}$ for all three sectors
[the value of $N_b$ depends on the ratio $R/(\sqrt{\mu}r_0)$].
We note that
the integration scheme employed has not been 
optimized carefully and can possibly be refined further.
For the five-body system,
appropriate generalizations are introduced.

To demonstrate that the developed framework works, we consider 
four-body systems consisting of two identical spin-up fermions
and two identical spin-down fermions. Such systems can be realized 
experimentally by
occupying two different hyperfine states of
ultracold atomic $^6$Li or $^{40}$K 
samples~\cite{giorginireview,blochreview,doertereview}.
We assume that the identical fermions do not interact.
This is a good assumption since $s$-wave interactions
between identical fermions 
are forbidden by the Pauli exclusion principle
and $p$-wave interactions are, in most experimental
realizations, highly suppressed by the threshold law.
Furthermore, we assume that the interspecies interactions have
been tuned so that the interspecies
free-space $s$-wave scattering length
$a_s$ is infinitely large.
This regime is referred to as the unitary regime and
can be realized experimentally by applying an external magnetic
field in the vicinity of a Fano-Feshbach resonance~\cite{chinreview}.
In our calculations, we describe the interspecies
two-body interactions using a Gaussian two-body potential
with depth $D$ and range $r_0$ adjusted such that the two-body
potential supports a single zero-energy $s$-wave bound state.
We calculate the energetically
lowest-lying hyperangular eigenvalue $U_0(R)$ for various
$\sqrt{\mu}r_0/R$ values.

To present our results, we rewrite $U_0(R)$
in terms of the quantity $s_0(R)$,
\begin{eqnarray}
\label{eq_s0}
U_0(R) = \frac{\hbar^2 \{ [s_0(R)]^2-1/4 \}}{2 R^2}.
\end{eqnarray}
The scaling introduced  in Eq.~(\ref{eq_s0})
is motivated by the fact that 
$s_0(R)$ becomes independent of $R$ in the non-interacting 
limit and if $r_0 \ll |a_s|$~\cite{werner}. 
The latter regime is realized if the
$s$-wave scattering length diverges and if the quantity
$\sqrt{\mu}r_0$ is much smaller than $R$.

\begin{figure}
\vspace*{+1.cm}
\includegraphics[angle=0,width=70mm]{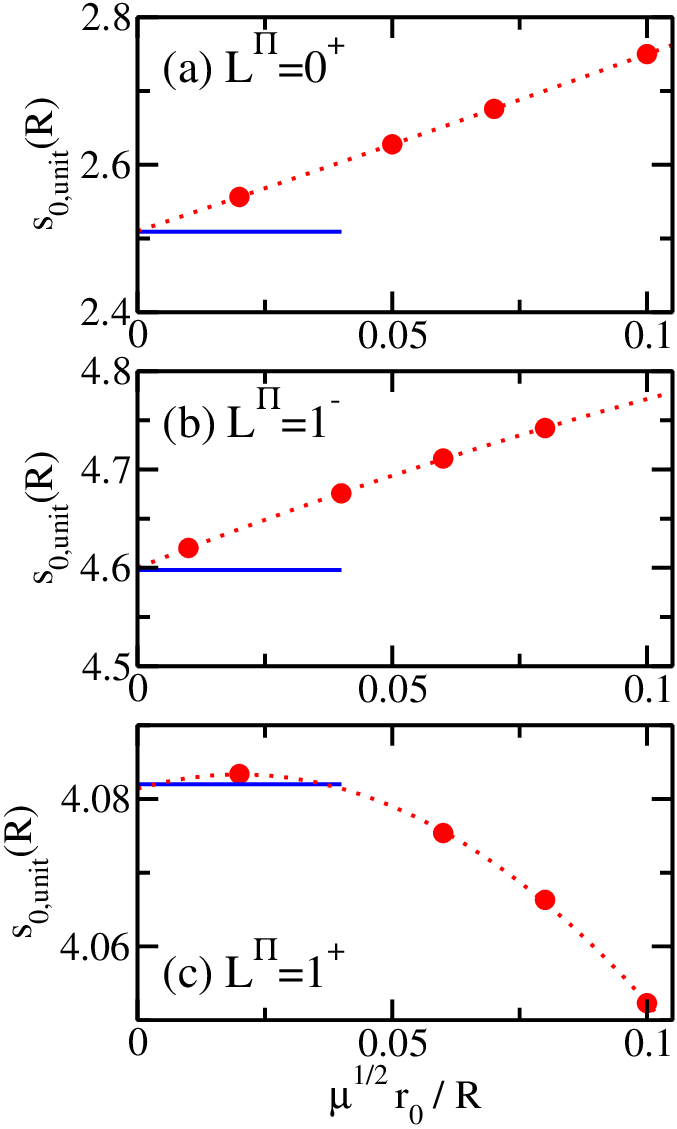}
\vspace*{1.cm}
\caption{(Color online)
Scaled hyperangular eigenvalue $s_{0,{\rm{unit}}}(R)$ for the $(2,2)$
system with $\kappa=1$ at unitarity.
Circles show the scaled hyperangular eigenvalue $s_{0,{\rm{unit}}}(R)$
as a function of $\sqrt{\mu}r_0/R$ 
for 
(a) $L^{\Pi}=0^+$ symmetry,
(b) $L^{\Pi}=1^-$ symmetry, and
(c) $L^{\Pi}=1^+$ symmetry.
The dotted lines show three-parameter fits (using a second-order 
polynomial).
The solid horizontal line in panel~(a) shows the result 
from Ref.~\cite{vonstecherPRA} for
$s_{0,{\rm{unit}}}^{\rm{ZR}}$ while
the solid horizontal lines in panels~(b) 
and (c) show the results 
from Ref.~\cite{debraj} for
$s_{0,{\rm{unit}}}^{\rm{ZR}}$
(see the text for details). 
The agreement between our results and those
from the literature is very good.
}\label{fig1}
\end{figure}

Circles in Fig.~\ref{fig1} show the quantity 
$s_{0,{\rm{unit}}}(R)$ for the $(2,2)$ system at unitarity as a 
function of $\sqrt{\mu}r_0/R$
for (a) $L^{\Pi}=0^+$ symmetry, 
(b) $L^{\Pi}=1^-$ symmetry, and
(c) $L^{\Pi}=1^+$ symmetry.
The basis set extrapolation error is smaller 
than the symbol size.
Dotted lines show a fit to the data using the fitting
function $s_{0,{\rm{unit}}}^{\rm{ZR}}+c_1x+c_2x^2$, where 
$x=\sqrt{\mu}r_0/R$.
The coefficient $s_{0,\rm{unit}}^{\rm{ZR}}$ is found to be
$2.510(1)$,
$4.600(3)$, and
$4.081(3)$ 
for $L^{\Pi}=0^+$, $1^-$ and $1^+$, respectively.
For the $(2,2)$ system with $0^+$ symmetry,
our results compare well with the 
value $2.5092$ obtained by von Stecher and Greene~\cite{vonstecherPRA}
using the same approach as that employed here.
For the $(2,2)$
systems with $1^-$ and $1^+$ symmetry, the $s_{0,{\rm{unit}}}^{\rm{ZR}}$ 
value has, to the best of
our knowledge, not been calculated within the hyperspherical
coordinate approach.
However, using scale invariance arguments~\cite{werner},
$s_{0,{\rm{unit}}}^{\rm{ZR}}$ can be extracted from the
energy spectrum of the harmonically trapped $(2,2)$
system. 
The $s_{0,\rm{unit}}^{\rm{ZR}}$ 
values for the zero-range system at unitarity,
obtained by analyzing the energy spectrum of the trapped system,
are 
$s_{0,\rm{unit}}^{\rm{ZR}}=4.5978$ 
for $L^{\Pi}=1^-$ symmetry 
and 
$s_{0,\rm{unit}}^{\rm{ZR}}=4.0820$ 
for $L^{\Pi}=1^+$ symmetry~\cite{debraj}. 
Our values reported above are in very good agreement with these
literature
values,
indicating that the HECG approach is capable of reliably describing 
strongly-correlated few-body systems with finite angular momentum
and positive and negative parity.

\begin{figure}
\vspace*{+1.cm}
\includegraphics[angle=0,width=70mm]{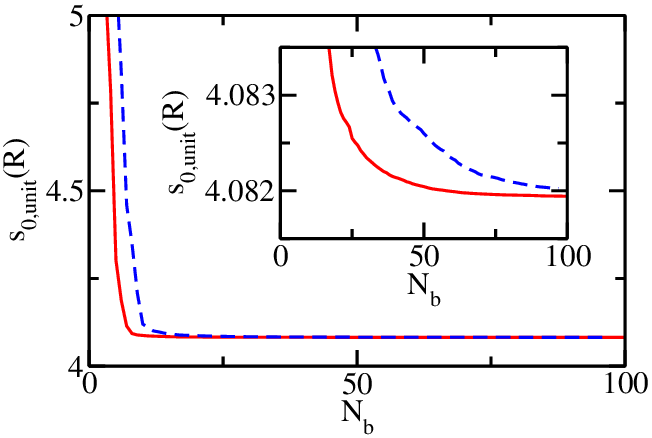}
\vspace*{1.cm}
\caption{(Color online)
Convergence of the scaled hyperangular eigenvalue
with increasing number of basis functions $N_b$.
Solid and dashed lines show the scaled hyperangular eigenvalue
$s_{0,{\rm{unit}}}(R)$ as a function of $N_b$ for the $(3,1)$
system with $1^+$ symmetry and $\kappa=1$
at unitarity for $\sqrt{\mu}r_0/R=0.02$
and $\sqrt{\mu}r_0/R=0.005$, respectively.
The inset shows a blowup.
}\label{fig2}
\end{figure}

To illustrate the convergence of the HECG approach with the number of 
basis functions, we consider the $(3,1)$ system with $1^+$ 
symmetry ($\kappa=1$ and $1/a_s=0$).
Solid and dashed lines 
in Fig.~\ref{fig2} show the scaled hyperangular eigenvalue
$s_{0,{\rm{unit}}}(R)$ as a function of the number of basis functions $N_b$ 
for $\sqrt{\mu}r_0/R=0.02$ and
$\sqrt{\mu}r_0/R=0.005$, respectively. For these calculations,
we considered $N_{\rm{T}}=4800$ and $6000$ trial functions, respectively,
for each new basis function selected.
Figure~\ref{fig2} shows that the description of the system becomes more
challenging as the separation of length scales increases,
i.e., as the ratio $\sqrt{\mu}r_0/R$ decreases.
Moreover, it can be seen that $s_{0,{\rm{unit}}}(R)$ shows a
few  ``shoulders'',
suggesting that there is room to improve upon the selection of the 
basis functions. Possible improvements may include gradient 
optimization techniques, which have been successfully employed in
electronic structure calculations~\cite{gradientoptimization}, or a refined
trial and error procedure.
Nevertheless, the HECG approach in its present implementation
yields good convergence 
for basis sets consisting of around 100 basis functions
for the systems under study.
The $s_{0,{\rm{unit}}}(R)$ for the largest basis set considered are
$4.08194$ and $4.0820$ for $\sqrt{\mu}r_0/R=0.02$ and
$\sqrt{\mu}r_0/R=0.005$, respectively.
The corresponding basis set errors are estimated to be
smaller than $0.00002$ and $0.0002$, respectively.
The dependence of $s_{0,{\rm{unit}}}(R)$ on $r_0$ is relatively weak
for the $(3,1)$ system with $1^+$ symmetry
and we find the extrapolated value
$s_{0,{\rm{unit}}}^{\rm{ZR}}=4.0820(3)$.
This 
result agrees well with the value of $s_{0,{\rm{unit}}}^{\rm{ZR}}=4.0819$
extracted from the trap energies~\cite{debraj}.

\begin{figure}
\vspace*{+1.cm}
\includegraphics[angle=0,width=70mm]{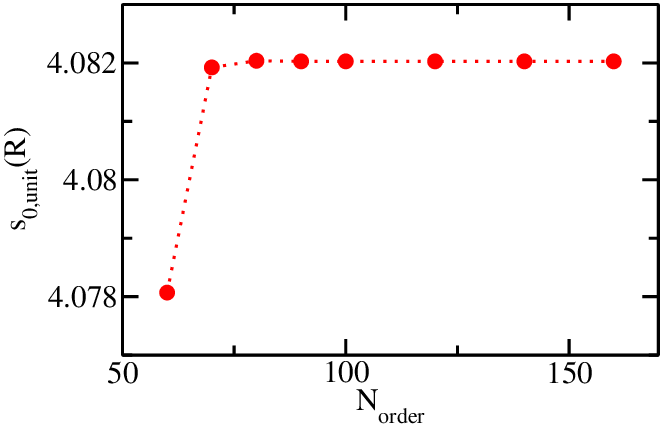}
\vspace*{1.cm}
\caption{(Color online)
Convergence of the scaled hyperangular eigenvalue with
increasing number of grid points.
The circles show the scaled hyperangular eigenvalue
$s_{0,{\rm{unit}}}(R)$, calculated for a basis set consisting of $N_b=95$
basis functions, as a function of the order $N_{\rm{order}}$
per sector
used to perform the numerical integration over the angle $\gamma_2$;
as discussed in the text, the numerical integration is divided
into three sectors.
For $N_{\rm{order}}=50$ (not shown), the numerical integration breaks
down
(it yields a value that deviates by 10\% from the exact value).
The calculations are performed 
for the $(3,1)$ system with $1^+$
symmetry, $1/a_s=0$, $\kappa=1$ and 
$\sqrt{\mu}r_0/R=0.005$.
The dotted line is shown as a guide to the eye.
}\label{fig3}
\end{figure}

Circles in Fig.~\ref{fig3} shows the scaled hyperangular eigenvalue
$s_{0,{\rm{unit}}}(R)$ for the $(3,1)$ system with $\sqrt{\mu}r_0/R=0.005$
and $1^+$ symmetry ($\kappa=1$ and $1/a_s=0$) as a function of the order
$N_{\rm{order}}$ used to perform the numerical integration over the
hyperangle $\gamma_2$. As discussed above, the numerical integral
is divided dynamically into three sectors, yielding a total of
$3 N_{\rm{order}}$ integration points.
For the calculations shown in Fig.~\ref{fig3}, we used 
a basis set with $N_b=95$. Figure~\ref{fig3} shows that the
scaled hyperangular eigenvalue $s_{0,{\rm{unit}}}(R)$ is, 
for the system considered,
accurate to better than 0.1\% for $N_{\rm{order}} \gtrsim 60$.
It is important to note, though, that while $N_{\rm{order}}=60$
yields quite accurate results, $N_{\rm{order}}=50$ yields
completely unreliable results for the parameter combination and basis
set considered.
In practice, we perform the optimization of the basis set for fixed
$N_{\rm{order}}$. At the end of the 
construction of the basis set, we increase
$N_{\rm{order}}$ to ensure that the results are independent
of the integration scheme employed.
In general, the smaller the ratio $\sqrt{\mu}r_0/R$, the larger
$N_{\rm{order}}$ (assuming the same $L$, $\Pi$, $a_s$ and $\kappa$).

As pointed out in 
Sec.~\ref{sec_matrixelements}, 
the
expressions for the matrix elements presented
in the appendices
apply to any number of particles.
To explicitly confirm this, we performed calculations for the 
non-interacting five-body systems with $0^+$, $1^-$
and $1^+$
symmetry and found hyperangular eigenvalues  consistent with
what is expected.
Since the matrix elements for the two-body interactions are of the 
same form as those 
for the overlaps, treatment of the non-interacting systems
suffices for testing the analytical expressions presented in this paper.
The computational demands will, of course, increase as the interactions 
are turned on. Assessing the performance of the outlined 
formalism for strongly-correlated
five-body systems is beyond the scope of this paper.

\section{$(3,1)$ system with $1^+$ symmetry}
\label{sec_applications}
This section considers the $(3,1)$ system with $1^+$ symmetry
at unitarity for various mass ratios.
The $s_{0,{\rm{unit}}}^{\rm{ZR}}$ value for these systems has been 
determined previously~\cite{blume1,blume2} by investigating the $(3,1)$
system under spherically symmetric harmonic confinement
using the stochastic variational approach combined
with geminal type basis functions, which are neither
characterized by 
good angular momentum and corresponding projection
quantum numbers 
nor good parity.
As a result, the earlier calculations were restricted to
comparatively large $r_0/a_{\rm{ho}}$ values, where
$a_{\rm{ho}}$ denotes the harmonic oscillator length
of the external confinement.
The present work determines $s_{0,{\rm{unit}}}^{\rm{ZR}}$ for various
mass ratios
$\kappa$ by employing the HECG approach. For comparative purposes,
we repeat the trap calculations using the standard ECG approach; however,
instead of using geminal type basis functions we employ
basis functions which are characterized by
good $L$, $M_L$ and $\Pi$ quantum numbers,
thereby allowing us to reduce the basis set extrapolation error
and to treat systems with smaller $r_0/a_{\rm{ho}}$ than considered earlier.

\begin{figure}
\vspace*{+1.cm}
\includegraphics[angle=0,width=70mm]{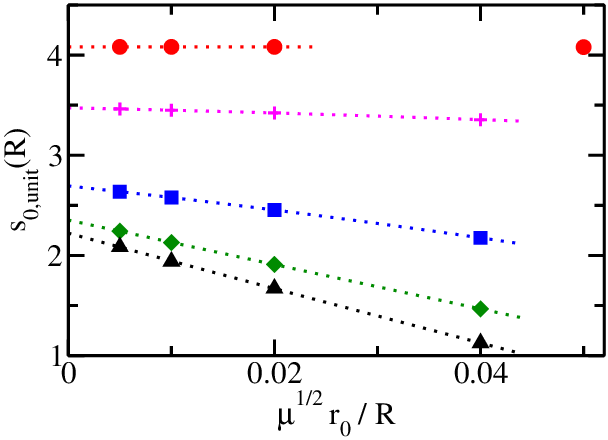}
\vspace*{1.cm}
\caption{(Color online)
$s_{0,{\rm{unit}}}(R)$ as a function of $\sqrt{\mu} r_0/R$
for the $(3,1)$ system with $1^+$ symmetry
at unitarity
for
$\kappa=1$ (circles),
$\kappa=4$ (pluses),
$\kappa=8$ (squares),
$\kappa=9.5$ (diamonds), and
$\kappa=10$ (triangles).
Dotted lines show three-parameter fits to the data.
}\label{fig4}
\end{figure}

Symbols in Fig.~\ref{fig4} show $s_{0,{\rm{unit}}}(R)$ 
[see Eq.~(\ref{eq_s0})], obtained by the HECG approach,
as a function of $\sqrt{\mu} r_0/R$ for
$\kappa=1, 4, 8, 9.5$ and $10$.
Dotted lines show three-parameter fits.
The resulting $s_{0,{\rm{unit}}}^{\rm{ZR}}$
values are summarized in column 2 of Table~\ref{tab1}.
\begin{table}
\caption{The first column shows the mass ratio $\kappa$.
Columns 2 and 3 show the $s_{0,{\rm{unit}}}^{\rm{ZR}}$
values for the $(3,1)$ system with $1^+$ symmetry at
unitarity obtained by the HECG approach and from the extrapolated
zero-range energies of the trapped system (see the text for details).
For comparison, column 4 shows the results from  Ref.~\cite{blume2}.
The $\kappa=1$ entry in the third column is taken from 
Ref.~\protect\cite{debraj}.
}
\begin{ruledtabular}
\begin{tabular}{cccc}
$\kappa$ & $s_{0,{\rm{unit}}}^{\rm{ZR}}$ & $s_{0,{\rm{unit}}}^{\rm{ZR}}$ & $s_{0,{\rm{unit}}}^{\rm{ZR}}$  \\
 & (HECG) & (trap, this work) & (trap, Ref.~\protect\cite{blume2})\\
\hline
$1/10$ &            & 4.4256(1) & \\
$1/5$  &            & 4.3663(1) & \\
$2/5$  &            & 4.2735(1) & \\
$3/5$  &            & 4.2000(1) & \\
$4/5$  &            & 4.1374(2) & \\
$1$    & 4.0820(3)  & 4.0819(1) & 4.08 \\
$2$    &            & 3.8532(5) & 3.86 \\
$3$    &            & 3.657(1)  &  \\
$4$    & 3.474(4)   & 3.472(2)  & 3.51 \\
$8$    & 2.69(3)    & 2.68(1)   & 2.79 \\
$9$    &            & 2.41(2)   & \\
$19/2$ & 2.35(6)    &           & \\
$10$   & 2.22(10)   &           & \\
\end{tabular}
\end{ruledtabular}
\label{tab1}
\end{table}
The errorbars are primarily due to the extrapolation 
to the $\sqrt{\mu}r_0/R \rightarrow 0$
limit and only secondarily due to the basis set
extrapolation error of the $s_{0,\rm{unit}}(R)$
for each $R$.
Figure~\ref{fig4} suggests that the leading order correction
to $s_{0,{\rm{unit}}}^{\rm{ZR}}$ is proportional to $1/R$
for the $\kappa$ values considered ($\kappa>1$).
This is in agreement with what has been found analytically 
for the $(2,1)$ system with $1^-$ symmetry~\cite{kart07}.
Figure~\ref{fig4} also shows that
the range dependence of $s_{0,{\rm{unit}}}(R)$ 
increases with increasing $\kappa$. For larger $\kappa$,
the range dependence appears to be more complicated
and the determination
of the corresponding $s_{0,{\rm{unit}}}(R)$ 
values is beyond the scope of this
paper.

\begin{figure}
\vspace*{+1.cm}
\includegraphics[angle=0,width=70mm]{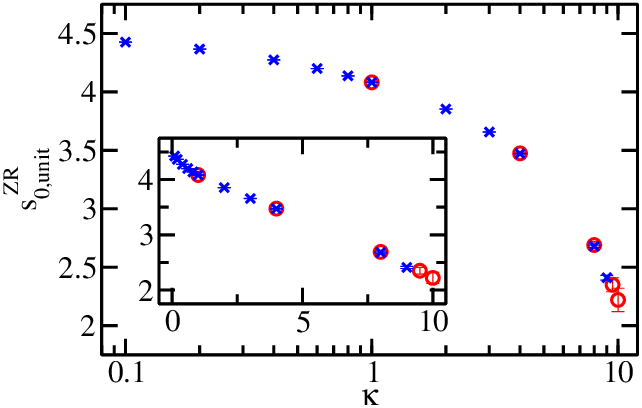}
\vspace*{1.cm}
\caption{(Color online)
$s_{0,{\rm{unit}}}^{\rm{ZR}}$ as a function of $\kappa$
for the $(3,1)$ system with $1^+$ symmetry
at unitarity.
The circles show the $s_{0,{\rm{unit}}}^{\rm{ZR}}$
values determined using the HECG approach while
the crosses show the $s_{0,{\rm{unit}}}^{\rm{ZR}}$
values determined by extrapolating the trap energies
obtained by the standard ECG approach to the zero-range
limit.
The errorbars increase with increasing $\kappa$.
The agreement between the $s_{0,{\rm{unit}}}^{\rm{ZR}}$
values determined by the two different approaches is very good.
The main panel and inset show the same data
on different scales; the main panel and inset
use respectively logarithmic and linear scales for $\kappa$.
}\label{fig5}
\end{figure}

Circles in Fig.~\ref{fig5}
show the $s_{0,{\rm{unit}}}^{\rm{ZR}}$
values obtained from the HECG approach
as a function of $\kappa$.
For comparison, crosses show the $s_{0,{\rm{unit}}}^{\rm{ZR}}$
values obtained 
by extrapolating the finite-range energies of the trapped system
to the $r_0/a_{\rm{ho}} \rightarrow 0$ limit.
These $s_{0,{\rm{unit}}}^{\rm{ZR}}$ values are reported in 
column 3 of Table~\ref{tab1} and are
calculated following the procedure discussed in
Ref.~\cite{debraj} for equal masses.
The agreement between the two sets of calculations is very good.
Column~4 of Table~\ref{tab1}
shows the $s_{0,{\rm{unit}}}^{\rm{ZR}}$ values obtained earlier~\cite{blume2};
these earlier calculations were restricted to larger $r_0/a_{\rm{ho}}$
and are 
less accurate than the values calculated in this
work.

Following the discussion of 
Refs.~\cite{petrov,werner,blume1,blume2,nishida},
the $s_{0,{\rm{unit}}}^{\rm{ZR}}$ value indicates whether the 
system behaves universal or not. For two-component
Fermi gases with zero-range $s$-wave interactions, e.g., the
$s_{0,{\rm{unit}}}^{\rm{ZR}}$ value is larger than 1 and, correspondingly,
the system properties are fully determined
by $a_s$.
If $s_{0,{\rm{unit}}}^{\rm{ZR}}<1$, the 
solution to the hyperradial Schr\"odinger equation,
which is a second order differential equation, 
can---at least in principle---contain contributions
of the ``regular'' and ``irregular'' solutions.
If the irregular solution contributes, the system is said 
to behave non-universal since its properties depend not 
only on the $s$-wave scattering length
but additionally on a second parameter.
Applying these arguments to the $(3,1)$ system
with $1^+$ symmetry 
and
using that $s_{0,{\rm{unit}}}^{\rm{ZR}}>1$ for the mass ratios considered,
the present study 
supports the finding that the four-body bound states 
found in Ref.~\cite{blum12} for $\kappa \gtrsim 9.5$ and positive
$s$-wave scattering length are
universal.

\section{Conclusions}
\label{sec_conclusions}
This paper extended the HECG approach, which had previously been
formulated for and applied to three- and four-particle systems
with $L^{\Pi}=0^+$ symmetry~\cite{vonstecherPRA,vonstecherThesis,ritt11}, 
to states with $1^-$ and $1^+$ symmetry.
The developed framework is applicable to systems with any $n$;
realistically, though, applications in the not too
distant future will likely be limited to 
systems with up to five particles.
This paper emphasized a unified formulation
for solving the hyperangular Schr\"odinger equation.
In particular, many of
the resulting equations apply to any particle number and 
$L^{\Pi}$ symmetry, suggesting 
a numerical implementation in
which most subroutines can be used for any 
particle number $n$ and any $L^{\Pi}$ symmetry;
only a few subroutines specific to the values of
$n$, $L$ and $\Pi$ are needed.

As a first application, we considered
the $(2,2)$ and $(3,1)$ systems at unitarity.
In particular, we solved the hyperangular 
Schr\"odinger equation for the energetically lowest-lying
eigenvalue in the 
small $\sqrt{\mu} r_0/R$ regime and 
extracted the corresponding $s_{0,{\rm{unit}}}^{\rm{ZR}}$ values.
Our results are in very good agreement with results from the literature
and with results determined by an alternative approach.
The $s_{0,{\rm{unit}}}^{\rm{ZR}}$ values
for the $(3,1)$ system at unitarity with $1^+$ symmetry consisting of
three heavy identical fermions and one light  impurity particle
are relevant to
the $(3,1)$ system with positive $a_s$. In particular,
the results obtained in this paper lend strong support that
the bound states of the $(3,1)$ system with positive 
$s$-wave scattering length found in Ref.~\cite{blum12} are 
universal, i.e., fully determined by $a_s$.
While we were able to reliably describe
four-body systems for which the hyperradius $R$ is 200 times larger
than the scaled range $\sqrt{\mu} r_0$ of the two-body
potential, pushing this ratio to much larger values may be challenging.

In the future, it will be interesting to combine,
as done in Refs.~\cite{vonstecherPRA,vonstecherThesis,ritt11},  the developed
framework with a standard R-matrix approach and to describe
the scattering properties of four-particle systems with 
finite angular momentum. The
generalization of the developed formalism to states with 
other $L^{\Pi}$ symmetries, which amounts
to determining the corresponding $d$-, $b$-,
$p$- and $q$-coefficients, 
is
tedious but straightforward. Other possible 
extensions include the generalization
of the approach to cold atom systems in a wave guide geometry.

\appendix

\section{Definition of (symmetry-dependent) auxiliary quantities}
\label{appendix_symmetry}
We first introduce a number of 
auxiliary quantities 
and then define quantities specific to the basis functions with 
$L^{\Pi}=0^+,1^-$ and $1^+$ symmetry.

The matrix $\underline{S}$ is defined as
\begin{eqnarray}
\underline{S} = 
\underline{A}_B \underline{A}_B'+
\underline{A}_B' \underline{A}_B,
\end{eqnarray}
where
\begin{eqnarray}
\underline{A}_B = (\underline{U}_B)^T \underline{A}
\underline{U}_B
\end{eqnarray}
and
\begin{eqnarray}
\underline{A}_B' = (\underline{U}_B)^T \underline{A}'
\underline{U}_B.
\end{eqnarray}
Similarly, we define
\begin{eqnarray}
\underline{B}_B =\underline{A}_B+\underline{A}'_B.
\end{eqnarray}
We define the 
vectors
$\vec{u}_{j,B}$ and
$\vec{u}_{j,B}'$
 ($j=1$ and 2),
\begin{eqnarray}
\vec{u}_{j,B} = \left( \underline{U}_B \right)^T \vec{u}_j
\end{eqnarray}
and
\begin{eqnarray}
\vec{u}_{j,B}' = \left( \underline{U}_B \right)^T \vec{u}_j' .
\end{eqnarray}
Lastly, we define
\begin{eqnarray}
\label{eq_definition_ajk}
a(j,k) = u_{1,B}(j) u_{1,B}'(k),
\end{eqnarray}
\begin{eqnarray}
\bar{a}(j,k) = u_{2,B}(j) u_{2,B}'(k),
\end{eqnarray}
\begin{eqnarray}
g(j,k,l)=
a(j,j) \bar{a}(l,k)+
a(l,k) \bar{a}(j,j)- \nonumber \\
a(l,j) \bar{a}(j,k)-
a(j,k) \bar{a}(l,j)
,
\end{eqnarray}
\begin{eqnarray}
\bar{g}(j,k,l)=g(j,l,k),
\end{eqnarray}
\begin{eqnarray}
h(j,k,l)=g(l,j,k)+\bar{g}(l,j,k),
\end{eqnarray}
\begin{eqnarray}
F(j,k,l,m)=
a(j,k)\bar{a}(l,m)
+a(k,j)\bar{a}(l,m)
+ \nonumber \\
a(j,k)\bar{a}(m,l)
+a(k,j)\bar{a}(m,l)
,
\end{eqnarray}
\begin{eqnarray}
G(j,k,l,m)=
a(j,l)\bar{a}(m,k)
+a(j,m)\bar{a}(l,k)
+ \nonumber \\
a(k,l)\bar{a}(m,j)
+a(k,m)\bar{a}(l,j)
,
\end{eqnarray}
and
\begin{eqnarray}
f(j,k,l,m)=
F(j,k,l,m)+
F(l,m,j,k)
- \nonumber \\
G(j,k,l,m)
-G(l,m,j,k)
.
\end{eqnarray}

Sections~\ref{appendixc_zeroplus}-\ref{appendixc_oneplus}
give explicit expressions for the $d$-, $p$-, $q$- and
$b$-coefficients for the basis functions with $0^+$, $1^-$ and $1^+$
symmetry, respectively.
In what follows, the elements of the vector $\vec{u}_{j,B}$ are denoted by
${u}_{j,B}(k)$, $k=1,\cdots,n-1$; the same
notation is adopted for the elements of 
other
vectors.
The elements of the matrix $\underline{S}$ are denoted by
$\underline{S}(k,l)$ with $k$ and $l=1,\cdots,n-1$;
the same notation is adopted for the elements of other
matrices.

\subsection{$0^+$ symmetry}
\label{appendixc_zeroplus}
The only non-zero $d$-coefficient is $d^{(0)}$,
\begin{eqnarray}
d^{(0)}=1.
\end{eqnarray}

The only non-zero $p$-coefficient is
$p_j^{(2)}$,
\begin{eqnarray}
p_j^{(2)}=\underline{A}_B(j,j) R^{-1}.
\end{eqnarray}

The non-zero $q$-coefficients are
$q_j^{(2)}$, $q_j^{(4)}$ and $q_{j,k}^{(22)}$,
\begin{eqnarray}
q_j^{(2)} = 
R^{-1} p_j^{(2)},
\end{eqnarray}
\begin{eqnarray}
q_j^{(4)} = -\underline{A}_B^2(j,j) R^{-2},
\end{eqnarray}
and
\begin{eqnarray}
q_{j,k}^{(22)} = -\left[
2\underline{A}_B(j,j) \underline{A}_B(k,k) 
+
\frac{4}{3} \underline{A}_B^2(j,k) 
\right]
R^{-2}.
\end{eqnarray}
Here, $\underline{A}_B^2(j,k)$ denotes the 
square of the matrix element
$\underline{A}_B(j,k)$.

The non-zero $b$-coefficients are $b^{(0)}$, $b_j^{(2)}$,
$b_j^{(4)}$ and $b^{(22)}_{j,k}$,
\begin{eqnarray}
b^{(0)} = -3 
\mbox{Tr}(\underline{B}_B),
\end{eqnarray}
\begin{eqnarray}
b_j^{(2)} = \beta_j^2 - \underline{S}(j,j)+ 
(3n-3) 
\beta_j R^{-2}  ,
\end{eqnarray}
\begin{eqnarray}
b_j^{(4)} = 
\left[-\beta_j^2 
+ 2 
\underline{A}_B(j,j) \underline{A}_B'(j,j)
\right] R^{-2} ,
\end{eqnarray}
and
\begin{eqnarray}
b_{j,k}^{(22)} = 
-2\beta_j \beta_k R^{-2} + 
2\underline{A}_B(j,j) \underline{A}_B'(k,k)R^{-2} +
\nonumber \\
2\underline{A}_B(k,k) \underline{A}_B'(j,j)R^{-2} +
\frac{8}{3} \underline{A}_B(j,k) \underline{A}_B'(j,k)
R^{-2} .
\end{eqnarray}

\subsection{$1^-$ symmetry}
\label{appendixc_oneminus}
The only non-zero $d$-coefficient is $d^{(2)}_j$,
\begin{eqnarray}
d^{(2)}_j=a(j,j).
\end{eqnarray}

The non-zero $p$-coefficients are
$p_j^{(2)}$, $p_j^{(4)}$ and $p_{j,k}^{(22)}$,
\begin{eqnarray}
p_j^{(2)} = -a(j,j) R^{-1},
\end{eqnarray}
\begin{eqnarray}
p_j^{(4)} = \underline{A}_B(j,j) a(j,j) R^{-1},
\end{eqnarray}
and
\begin{eqnarray}
p_{j,k}^{(22)} = 
\underline{A}_B(j,j) a(k,k) R^{-1} + \nonumber \\
\underline{A}_B(k,k) a(j,j) R^{-1} + \nonumber \\
\frac{2}{3} \underline{A}_B(j,k) 
\left[ a(j,k) + a(k,j) \right] R^{-1}.
\end{eqnarray}

The non-zero $q$-coefficients are
$q_j^{(4)}$, $q_{j,k}^{(22)}$, $q_j^{(6)}$, $q_{j,k}^{(24)}$ 
and $q_{j,k,l}^{(222)}$,
\begin{eqnarray}
q_j^{(4)} = 3 R^{-1} p_j^{(4)},
\end{eqnarray}
\begin{eqnarray}
q_{j,k}^{(22)} = 3 R^{-1} p_{j,k}^{(22)},
\end{eqnarray}
\begin{eqnarray}
q_j^{(6)} = - \underline{A}_B^2(j,j) a(j,j) R^{-2},
\end{eqnarray}
\begin{eqnarray}
q_{j,k}^{(24)} = 
-\underline{A}_B^2(k,k) a(j,j)R^{-2}- \nonumber \\
 2 \underline{A}_B(j,j) \underline{A}_B(k,k) a(k,k)R^{-2} 
- \nonumber \\
\frac{4}{3} \underline{A}_B^2(j,k) a(k,k)R^{-2} -
\nonumber
\\
\frac{4}{3} \underline{A}_B(k,k) \underline{A}_B(j,k) 
\left[
a(j,k) + a(k,j) \right] R^{-2}
\end{eqnarray}
and
\begin{eqnarray}
q_{j,k,l}^{(222)} = \nonumber \\
\left[
-2 \underline{A}_B(j,j) \underline{A}_B(k,k) a(l,l)
-\frac{4}{3} \underline{A}_B^2(j,k) a(l,l) \right] R^{-2} +
\nonumber
\\
\left[ -\frac{4}{3} \underline{A}_B(j,k) \underline{A}_B(l,l)
-
\frac{8}{9} \underline{A}_B(j,l) \underline{A}_B(k,l) \right]
\times \nonumber \\
\left[
a(j,k) + a(k,j)  \right] R^{-2}.
\end{eqnarray}

The non-zero $b$-coefficients are
$b_j^{(2)}$, $b_j^{(4)}$, $b_j^{(6)}$,
$b_{j,k}^{(22)}$, $b_{j,k}^{(24)}$ and $b_{j,k,l}^{(222)}$,
\begin{eqnarray}
b^{(2)}_j=
-2 (3n-4) a(j,j) R^{-2}
- \nonumber \\
2\sum_{k=1}^{n-1} 
\left[ 
\underline{A}_B(j,k) a(k,j)+\underline{A}'_B(j,k) a(j, k) 
\right]
-
\nonumber \\
3 \mbox{Tr}(\underline{B}_B) a(j,j),
\end{eqnarray}
\begin{eqnarray}
b^{(4)}_j=
\left[ \beta_{j}^2 - \underline{S}(j,j) 
+(3n-1)\beta_j R^{-2} 
\right] a(j,j) 
,
\end{eqnarray}
\begin{eqnarray}
b^{(6)}_j=
\left[
-\beta_{j}^2+ 2 \underline{A}_{B}(j,j) \underline{A}_{B}'(j,j)
\right] 
a(j,j) R^{-2} ,
\end{eqnarray}
\begin{eqnarray}
b^{(22)}_{j,k}
=&&
\left[ 
\beta_{j}^2 - \underline{S}(j,j) 
+ (3n-1) \beta_j R^{-2} 
\right] 
a(k,k) +
\nonumber \\
&&
\left[ 
\beta_{k}^2 - \underline{S}(k,k) 
+ (3n-1) \beta_k R^{-2} 
\right] 
a(j,j) 
- \nonumber \\
&&\frac{1}{3} \left[ \underline{S}(j,k)+ \underline{S}(k,j) \right] 
\left[ a(j,k)+a(k,j) \right],
\end{eqnarray}
\begin{eqnarray}
b_{j,k}^{(24)}=
&& 
\left[
-\beta_k^2
+ 2 \underline{A}_B(k,k) \underline{A}_B'(k,k) 
\right]
a(j,j) R^{-2}+ 
\nonumber \\
&&
\big[
-2\beta_j \beta_k 
+2 \underline{A}_B(j,j) \underline{A}'_B(k,k)  
+
\nonumber \\
&&2 \underline{A}_B(k,k) \underline{A}'_B(j,j)
+ \frac{8}{3} \underline{A}_B(j,k) \underline{A}_B'(j,k)
\big] \times \nonumber \\  
&&a(k,k) R^{-2} +\nonumber \\
&&
\frac{4}{3} 
\left[
\underline{A}_B(j,k) \underline{A}_B'(k,k)
+\underline{A}_B(k,k) \underline{A}_B'(j,k)
\right] \times \nonumber \\
&&\left[
a(j,k)+a(k,j)
\right]
R^{-2}
\end{eqnarray}
and
\begin{eqnarray}
b_{j,k,l}^{(222)}=
&&
\big[
-2\beta_j \beta_k + 
2\underline{A}_B(j,j) \underline{A}_B'(k,k) + \nonumber \\
&&
2\underline{A}_B(k,k) \underline{A}_B'(j,j) +
\frac{8}{3} \underline{A}_B(j,k) \underline{A}'(j,k)
\big] \times \nonumber \\
&&a(l,l)  R^{-2} +
 \nonumber \\
&&
\frac{4}{3} 
\left[
\underline{A}_B(j,k) \underline{A}_B'(l,l)+
\underline{A}_B(l,l) \underline{A}_B'(j,k)
\right] \times \nonumber \\
&&\left[
a(j,k)+a(k,j)
\right] R^{-2} +
\nonumber \\
&&
\frac{8}{9}
\left[
\underline{A}_B(j,l) \underline{A}_B'(k,l)+
\underline{A}_B(k,l) \underline{A}_B'(j,l)
\right] \times
\nonumber \\
&&\left[
a(j,k) + a(k,j)
\right] R^{-2}
.
\end{eqnarray}

\subsection{$1^+$ symmetry}
\label{appendixc_oneplus}

The only non-zero $d$-coefficient is $d_{j,k}^{(22)}$,
\begin{eqnarray}
d_{j,k}^{(22)}=
a(j,j) \bar{a}(k,k) + a(k,k) \bar{a}(j,j) 
-
\nonumber \\
a(j,k) \bar{a}(k,j)- a(k,j) \bar{a}(j,k) .
\end{eqnarray}

The non-zero $p$-coefficients are $p_{j,k}^{(22)}$,
$p_{j,k}^{(24)}$ and $p_{j,k,l}^{(222)}$,
\begin{eqnarray}
p_{j,k}^{(22)} = -2R^{-1} d_{j,k}^{(22)},
\end{eqnarray}
\begin{eqnarray}
p_{j,k}^{(24)} = R^{-1} d_{j,k}^{(22)} \underline{A}_B(k,k),
\end{eqnarray}
and
\begin{eqnarray}
p_{j,k,l}^{(222)} = 
R^{-1} d_{j,k}^{(22)} \underline{A}_B(l,l) 
+ \nonumber \\
\frac{2}{3}R^{-1} \underline{A}_B(j,k) h(j,k,l).
\end{eqnarray}

The non-zero $q$-coefficients are
$q_{j,k}^{(22)}$, $q_{j,k}^{(24)}$, $q_{j,k,l}^{(222)}$,
$q_{j,k}^{(26)}$,
$q_{j,k}^{(44)}$,
$q_{j,k,l}^{(224)}$ and
$q_{j,k,l,m}^{(2222)}$,
\begin{eqnarray}
q_{j,k}^{(22)} = R^{-1} p_{j,k}^{(22)},
\end{eqnarray}
\begin{eqnarray}
q_{j,k}^{(24)} = 5 R^{-1} p_{j,k}^{(24)},
\end{eqnarray}
\begin{eqnarray}
q_{j,k,l}^{(222)} = 5 R^{-1} p_{j,k,l}^{(222)},
\end{eqnarray}
\begin{eqnarray}
q_{j,k}^{(26)}=
-d_{j,k}^{(22)} \underline{A}_B^2(k,k) R^{-2},
\end{eqnarray}
\begin{eqnarray}
q_{j,k}^{(44)}=
- 2 d_{j,k}^{(22)} \underline{A}_B(j,j) \underline{A}_B(k,k) R^{-2} - 
\nonumber \\
\frac{4}{5} d_{j,k}^{(22)} \underline{A}_B^2(j,k) R^{-2},
\end{eqnarray}
\begin{eqnarray}
q_{j,k,l}^{(224)} =
&&- \left[
\underline{A}_B(j,j) \underline{A}_B(l,l) 
+ \frac{2}{3} \underline{A}_B^2(j,l) \right]
2 R^{-2} d_{k,l}^{(22)} 
-\nonumber \\
&& \left[
\underline{A}_B(k,k) \underline{A}_B(l,l) 
+ \frac{2}{3} \underline{A}_B^2(k,l) \right]
2 R^{-2} d_{j,l}^{(22)} 
- \nonumber \\
&&
 \underline{A}_B^2(l,l) R^{-2} d_{j,k}^{(22)} -
\nonumber \\
&& \frac{4}{3} 
\big[ \underline{A}_B(j,k) \underline{A}_B(l,l) h (j,k,l) 
+ \nonumber \\
&&\underline{A}_B(j,l) \underline{A}_B(l,l) h (j,l,k) 
+ \nonumber \\
&&\underline{A}_B(k,l) \underline{A}_B(l,l) h (k,l,j) 
\big] R^{-2}
,
\end{eqnarray}
and
\begin{eqnarray}
q_{j,k,l,m}^{(2222)}=
\nonumber \\
-
 \left[
2 \underline{A}_B(l,l) \underline{A}_B(m,m) 
+ \frac{4}{3} 
\underline{A}_B^2(l,m) 
\right]  R^{-2} d_{jk}^{(22)} 
-
\nonumber \\
\left[
2
\underline{A}_B(j,j) \underline{A}_B(k,k)  
+ 
\frac{4}{3} 
\underline{A}_B^2(j,k)  
\right]  R^{-2} d_{lm}^{(22)} -\nonumber \\
\frac{4}{3} 
\big[ 
\underline{A}_B(j,k) \underline{A}_B(m,m) h(j,k,l) +
\nonumber \\
\underline{A}_B(j,k) \underline{A}_B(l,l) 
h(j,k,m) \big] 
R^{-2} -\nonumber \\
\frac{4}{3} 
\big[ 
\underline{A}_B(l,m) \underline{A}_B(k,k) 
h(l,m,j) +
\nonumber \\
\underline{A}_B(l,m) \underline{A}_B(j,j) 
h(l,m,k) \big] R^{-2} -\nonumber \\
\frac{8}{9} \big[
\underline{A}_B(l,k) \underline{A}_B(k,m)  
h(l,m,j) +
\nonumber \\
\underline{A}_B(l,j) \underline{A}_B(j,m) 
h(l,m,k) \big]
R^{-2}- \nonumber \\
\frac{8}{9} \big[
\underline{A}_B(j,m) \underline{A}_B(m,k) 
h(j,k,l) +
\nonumber \\
\underline{A}_B(j,l) \underline{A}_B(l,k) 
h(j,k,m) \big]
R^{-2} -\nonumber \\
\frac{8}{9}
\left[
\underline{A}_B(j,k) \underline{A}_B(l,m)
\right]f(j,k,l,m) R^{-2}
.
\end{eqnarray}

The non-zero $b$-coefficients are
\begin{eqnarray}
b_{j,k}^{(22)} &=& 
-5 \mbox{Tr}(\underline{B}_B)d_{j,k}^{(22)} -
4 
(3n-3) R^{-2} d_{j,k}^{(22)} - \nonumber \\
&& 2 \sum_{l\ne j,k}^{n-1}
\big[
\underline{A}_B(j, l) g(k,j,l) + \underline{A}_B'(j, l) \bar{g}(k,j,l) + 
\nonumber \\
&&\underline{A}_B(k, l) g(j,k,l) + \underline{A}_B'(k, l) \bar{g}(j,k,l) -
\nonumber \\
&& \beta_l d_{jk}^{(22)} 
\big],
\end{eqnarray}
\begin{eqnarray}
b_{j,k}^{(24)} =
\left[ \beta_k^2 - \underline{S}(k,k) \right] d_{j,k}^{(22)}
+\beta_k 
(3n+1) R^{-2} d_{j,k}^{(22)},
\end{eqnarray}
\begin{eqnarray}
b_{j,k,l}^{(222)} =
\left[ \beta_l^2 - \underline{S}(l,l) \right] d_{j,k}^{(22)} 
- \nonumber \\
\frac{2}{3} h(j,k,l) \underline{S}(j,k) 
+
\beta_l 
(3n+1) R^{-2} 
d_{j,k}^{(22)},
\end{eqnarray}
\begin{eqnarray}
b_{j,k}^{(44)}=
\big[
-\beta_j \beta_k + \underline{A}_B(j,j) \underline{A}_B'(k,k)
+
\nonumber \\
\underline{A}_B(k,k) \underline{A}_B'(j,j) + 
\frac{4}{5} 
\underline{A}_B(j,k) \underline{A}_B'(j,k)
\big] 2 R^{-2} d_{j,k}^{(22)},
\end{eqnarray}
\begin{eqnarray}
b_{j,k}^{(26)}=
\left[ -\beta_k^2 + 2 \underline{A}_B(k,k) \underline{A}'_B(k,k) \right]
R^{-2} d_{j,k}^{(22)}
,
\end{eqnarray}
\begin{eqnarray}
b_{j,k,l}^{(224)}=
\big[ -\beta_j \beta_l + \underline{A}_B(j,j) \underline{A}_B'(l,l) + 
\underline{A}_B(l,l) \underline{A}_B'(j,j) +
\nonumber \\
\frac{4}{3} \underline{A}_B(j,l) \underline{A}_B'(j,l)
\big]
2 R^{-2} d_{k,l}^{(22)} +
\nonumber \\
\big[ -\beta_k \beta_l + \underline{A}_B(k,k) \underline{A}_B'(l,l) + 
\underline{A}_B(l,l) \underline{A}_B'(k,k) +
\nonumber \\
\frac{4}{3} \underline{A}_B(k,l) \underline{A}_B'(k,l)
\big]
2 R^{-2} d_{j,l}^{(22)} +\nonumber \\
\left[ -\beta_l^2 + 2 \underline{A}_B(l,l) \underline{A}_B'(l,l) \right]
 R^{-2} d_{j,k}^{(22)} -\nonumber \\
\frac{4}{3} 
\left[
\underline{A}_B(j,k) \underline{A}_B(l,l) + \underline{A}_B'(j,k)
\underline{A}_B'(l,l)
\right] R^{-2} h(j,k,l)- \nonumber \\
\frac{4}{3} 
\left[
\underline{A}_B(j,l) \underline{A}_B(l,l) + \underline{A}_B'(j,l)
\underline{A}_B'(l,l)
\right] R^{-2} h(j,l,k) -\nonumber \\
\frac{4}{3} 
\left[
\underline{A}_B(k,l) \underline{A}_B(l,l) + \underline{A}_B'(k,l)
\underline{A}_B'(l,l)
\right] R^{-2} h(k,l,j) 
,
\end{eqnarray}
and
\begin{eqnarray}
b_{j,k,l,m}^{(2222)}= \nonumber \\
-
 \big\{
2 \underline{A}_B(l,l) \underline{A}_B(m,m) +
2 \underline{A}_B'(l,l) \underline{A}_B'(m,m)
+ \nonumber \\
\frac{4}{3} \left[ 
\underline{A}_B^2(l,m) +\underline{A}_B'^2(l,m) \right]
\big\}  R^{-2} d_{jk}^{(22)} -\nonumber \\
 \big\{
2
\underline{A}_B(j,j) \underline{A}_B(k,k) +
2
\underline{A}_B'(j,j) \underline{A}_B'(k,k) 
+ \nonumber \\
\frac{4}{3} 
\left[ 
\underline{A}_B^2(j,k) +\underline{A}_B'^2(j,k) \right]
\big\}  R^{-2} d_{lm}^{(22)} -\nonumber \\
\frac{4}{3} 
\left[ 
\underline{A}_B(j,k) \underline{A}_B(m,m) +
\underline{A}_B'(j,k) \underline{A}_B'(m,m) 
\right] \times 
\nonumber \\
h(j,k,l) R^{-2} - \nonumber \\
\frac{4}{3} 
\left[ 
\underline{A}_B(j,k) \underline{A}_B(l,l) +
\underline{A}_B'(j,k) \underline{A}_B'(l,l) 
\right] \times
\nonumber \\
h(j,k,m) R^{-2}- \nonumber \\
\frac{4}{3} 
\left[ 
\underline{A}_B(l,m) \underline{A}_B(k,k) +
\underline{A}_B'(l,m) \underline{A}_B'(k,k) 
\right] \times 
\nonumber \\
h(l,m,j) R^{-2} - \nonumber \\
\frac{4}{3} 
\left[ 
\underline{A}_B(l,m) \underline{A}_B(j,j) +
\underline{A}_B'(l,m) \underline{A}_B'(j,j) 
\right] \times 
\nonumber \\
h(l,m,k) R^{-2} - \nonumber \\
\frac{8}{9} \left[
\underline{A}_B(l,k) \underline{A}_B(k,m) +
\underline{A}'_B(l,k) \underline{A}'_B(k,m) \right] \times
\nonumber \\
h(l,m,j) R^{-2} - \nonumber \\
\frac{8}{9} \left[
\underline{A}_B(l,j) \underline{A}_B(j,m) +
\underline{A}'_B(l,j) \underline{A}'_B(j,m) \right] \times
\nonumber \\
h(l,m,k) R^{-2} - \nonumber \\
\frac{8}{9} \left[
\underline{A}_B(j,m) \underline{A}_B(m,k) +
\underline{A}'_B(j,m) \underline{A}'_B(m,k) \right] \times
\nonumber \\
h(j,k,l) R^{-2} -\nonumber \\
\frac{8}{9} \left[
\underline{A}_B(j,l) \underline{A}_B(l,k) +
\underline{A}'_B(j,l) \underline{A}'_B(l,k) \right]
h(j,k,m) R^{-2} - \nonumber \\
\frac{8}{9}
\left[
\underline{A}_B(j,k) \underline{A}_B(l,m)+
\underline{A}_B'(j,k) \underline{A}_B'(l,m)
\right]
\times
\nonumber \\
f(j,k,l,m) R^{-2}
.
\end{eqnarray}

\section{Sketch of derivation of 
matrix elements
for $n=3$ and $L^{\Pi}=1^-$ symmetry}
\label{appendix_derivation}

This appendix derives the overlap
matrix element for $n=3$ and $L^{\Pi}=1^-$.
To evaluate the overlap matrix element,
we write, using
Eq.~(\ref{eq_basis_oneminus}),
\begin{eqnarray}
\label{eq_waveproduct}
\psi(\underline{A}',\vec{u}_1',\vec{x})
\psi(\underline{A} ,\vec{u}_1 ,\vec{x})
=
3 v_{1,3}' v_{1,3} \exp\left(- 
\frac{\vec{x}^T \underline{B} \vec{x}}{2} \right),
\end{eqnarray}
where $\underline{B}$ is defined in Eq.~(\ref{eq_bmatrix}).
Using the transformation
$\vec{y} = \underline{U}_B^T \vec{x}$,
we find
\begin{eqnarray}
\label{eq_vproduct1}
v_{1,3}' v_{1,3}
= \nonumber \\
a(1,1) y_{1,3}^2 + \left[ a(1,2) + a(2,1) \right] y_{1,3} y_{2,3}
+a(2,2) y_{2,3}^2 ,
\end{eqnarray}
where 
$a(i,j)$ is defined in Eq.~(\ref{eq_definition_ajk})
and where $y_{j,3}$ denotes the $z$-component
of the vector $\vec{y}_j$.
Using $y_{1,3}= y_1 \cos \vartheta_1$ and 
$y_{2,3}= y_2 \cos \vartheta_2$
in Eqs.~(\ref{eq_waveproduct}) and (\ref{eq_vproduct1}),
we have
\begin{eqnarray}
\psi'\psi =
3 
\big\{
a(1,1) y_1^2 \cos^2 \vartheta_1
+ \nonumber \\
\left[a(1,2)+a(2,1)\right]
y_1 y_2 \cos \vartheta_1 \cos \vartheta_2
+
a(2,2) y_2^2 \cos^2 \vartheta_2
\big\} 
\times \nonumber \\
\exp \left[ -\frac{1}{2} 
\left(\beta_1 y_1^2 +
\beta_2 y_2^2 \right) \right],
\end{eqnarray}
where the $\beta_j$ denote, as
before, the eigenvalues of the matrix $\underline{B}$.
When integrating 
over $\hat{y}_1 \hat{y}_2$, the cross term averages to zero
and we have
\begin{eqnarray}
\label{eq_wavepartialint}
\int_0^{2 \pi} 
\int_{-1}^{1} 
\int_0^{2 \pi} 
\int_{-1}^{1} 
\left( \psi'  \psi  \right)|_R
d \cos \vartheta_1 d \varphi_1 d \cos \vartheta_2 d \varphi_2
=\nonumber \\
(4 \pi)^2 
\left[
a(1,1) y_1^2
+
a(2,2) y_2^2
\right] \times \nonumber \\
\exp \left[ -\frac{1}{2} 
\left(\beta_1 y_1^2 +
\beta_2 y_2^2 \right) \right].
\end{eqnarray}
Comparison of  Eq.~(\ref{eq_wavepartialint})
with Eq.~(\ref{eq_overlap_firstint}) 
shows that 
$d_j^{(2)}=a(j,j)$
and that all other $d$-coefficients are zero.
This agrees with the expressions given in 
Appendix~\ref{appendixc_oneminus}.

Next, we consider the integration 
over $\gamma_1$. 
To this end, we replace $y_1$ and $y_2$ in Eq.~(\ref{eq_wavepartialint})
by $R \sin \gamma_1$ and $R \cos \gamma_1$, respectively.
According to the discussion in Sec.~\ref{sec_matrixelements},
$\gamma_1$ can take values
between $0$ and $\pi/2$.
Multiplying both sides of
Eq.~(\ref{eq_wavepartialint}) by
$\sin^2 \gamma_1 \cos^2 \gamma_1$ 
[see Eq.~(\ref{eq_hyperangularvolumeelement})]
and integrating over $\gamma_1$,
we find
\begin{eqnarray}
\label{eq_oneminusovfinal}
\int \left( \psi' \psi \right)|_R  d^5 \vec{\Omega}
= (4 \pi)^2 
\frac{\pi}{16 \zeta}
\exp\left[
-\frac{1}{4}R^2 (\beta_1+\beta_2)
\right] \times \nonumber \\
\big\{
\left[a(1,1)+a(2,2) \right] R^2 I_1(\zeta)
+ \nonumber \\
\left[-a(1,1)+a(2,2) \right] R^2 I_2(\zeta)
\big\}.
\end{eqnarray}
Applying the definitions from Appendix~\ref{appendix_symmetry}
and \ref{appendix_general}, it can be verified that 
Eq.~(\ref{eq_oneminusovfinal}) agrees with 
Eq.~(\ref{eq_integralovergammaone})
[note that Eq.~(\ref{eq_integralovergammaone}) 
does
not contain a factor of $(4 \pi)^2$ while
Eq.~(\ref{eq_oneminusovfinal}) does;
the reason is that
$f_o$ is defined without this prefactor].

The derivation sketched above for the overlap matrix element
can be fairly straightforwardly generalized to arbitrary $n$.
To calculate the matrix element 
$\langle \psi | T_{\Omega} | \psi \rangle |_R$,
we use Eq.~(\ref{eq_tomegasplit}),
i.e., we separately calculate the matrix
elements 
$\langle \psi | T_{\rm{rel}} | \psi \rangle |_R$ and
$\langle \psi | T_R | \psi \rangle |_R$. The evaluation of these matrix
elements is not fundamentally difficult but somewhat tedious and lengthy.

\section{Definitions of
$sc$-coefficients, $M_1$ and $M_2$}
\label{appendix_general}

The $sc$-coefficients entering into $f_o$
are given by
\begin{eqnarray}
\label{eq_scfirst}
sc_{00}&=&
d^{(0)}+\sum_{j=3}^{n-1} 
\left[d^{(2)}_j y_{j}^2+d^{(4)}_j y_{j}^4+d^{(6)}_j y_{j}^6
\right]+
\nonumber
\\
&&
\sum _{k>j;j,k \ne 1,2}^{n-1} \left[
d^{(22)}_{j,k} y_{j}^2y_{k}^2+ d^{(44)}_{j,k} y_j^4 y_k^4 \right] +
\nonumber
\\
&&
\sum _{j=3,k=3,k\neq j}^{n-1} 
\left[ d^{(24)}_{j,k} y_{j}^2 y_{k}^4 +
d^{(26)}_{j,k} y_j^2 y_k^6 \right] +
\nonumber
\\
&&
\sum _{k>j;l \ne j,k;j,k,l \ne 1,2}^{n-1} \left[
d^{(222)}_{j,k,l}y_{j}^2y_{k}^2y_{l}^2 +
d^{(224)}_{j, k,l} y_j^2 y_k^2 y_l^4 \right]+
\nonumber \\
&&\sum_{k>j;l>j;m>l;j \ne k \ne l \ne m;j,k,l,m \ne 1,2} ^{n-1}
d_{j,k,l,m}^{(2222)} y_j^2 y_k^2 k_l^2 y_m^2
,
\end{eqnarray}
\begin{eqnarray}
sc_{20}&=&
d^{(2)}_1+\sum_{j=3}^{n-1} 
\left[d_{1,j}^{(22)} y_{j}^2+d^{(24)}_{1,j} y_{j}^4+d^{(26)}_{1,j} y_{j}^6
\right]+
\nonumber
\\
&&
\sum _{j,k=3;k \ne j}^{n-1} \left[
d^{(222)}_{1,j,k} y_{j}^2y_{k}^2+ d^{(224)}_{1,j,k} y_j^2 y_k^4 \right] + 
\nonumber
\\
&&\sum _{j<k;j,k\neq 1,2}^{n-1} d^{(222)}_{j,k,1} y_{j}^2 y_{k}^2 +
\nonumber
\\
&&
\sum_{l>k;j \ne k \ne l;j,k,l \ne 1,2} ^{n-1}
d_{1,j,k,l}^{(2222)} y_j^2 y_k^2 k_l^2 
,
\end{eqnarray}
\begin{eqnarray}
sc_{02}&=&
d^{(2)}_2+\sum_{j=3}^{n-1} 
\left[d_{2,j}^{(22)} y_{j}^2+d^{(24)}_{2,j} y_{j}^4+d^{(26)}_{2,j} y_{j}^6
\right]+
\nonumber
\\
&&
\sum _{j,k=3;k \ne j}^{n-1} \left[
d^{(222)}_{2,j,k} y_{j}^2y_{k}^2+ d^{(224)}_{2,j,k} y_j^2 y_k^4 \right] + 
\nonumber
\\
&&\sum _{j<k;j,k\neq 1,2}^{n-1} d^{(222)}_{j,k,2} y_{j}^2 y_{k}^2 +
\nonumber
\\
&&
\sum_{l>k;j \ne k \ne l;j,k,l \ne 1,2} ^{n-1}
d_{2,j,k,l}^{(2222)} y_j^2 y_k^2 y_l^2 
,
\end{eqnarray}
\begin{eqnarray}
sc_{40}&=& d_1^{(4)} +
\sum_{j=3}^{n-1} \left[ d_{1,j}^{(44)} y_j^4
+                d_{j,1}^{(24)} y_j^2 \right] +
\nonumber
\\
&&
\sum_{j<k;j,k\ne 1,2}^{n-1} d_{j,k,1}^{(224)} y_j^2 y_k^2,
\end{eqnarray}
\begin{eqnarray}
sc_{04}&=& d_2^{(4)} +
\sum_{j=3}^{n-1} \left[ d_{2,j}^{(44)} y_j^4
+                       d_{j,2}^{(24)} y_j^2 \right] +
\nonumber
\\
&&
\sum_{j<k;j,k\ne 1,2}^{n-1} d_{j,k,2}^{(224)} y_j^2 y_k^2,
\end{eqnarray}
\begin{eqnarray}
sc_{60}&=& d_1^{(6)} + \sum_{j=3}^{n-1} d_{j,1}^{(26)} y_j^2,
\end{eqnarray}
\begin{eqnarray}
sc_{06}&=&d_2^{(6)} + \sum_{j=3}^{n-1} d_{j,2}^{(26)} y_j^2,
\end{eqnarray}
\begin{eqnarray}
sc_{22}&=& d_{1,2}^{(22)} +
\nonumber
\\
&&
\sum_{j=3}^{n-1} \left\{
\left[ 
d_{1,2,j}^{(222)} +d_{1,j,2}^{(222)} +d_{2,j,1}^{(222)} \right] y_j^2
+d_{1,2,j}^{(224)} y_j^4 
\right\} +
\nonumber
\\
&&
\sum_{j<k;j,k\ne 1,2}^{n-1} d_{12jk}^{(2222)} y_j^2 y_k^2
+
\nonumber
\\
&&
\sum_{j \ne k;j,k\ne 1,2}^{n-1} d_{1j2k}^{(2222)} y_j^2 y_k^2
,
\end{eqnarray}
\begin{eqnarray}
sc_{44}&=&d_{1,2}^{(44)},
\end{eqnarray}
\begin{eqnarray}
sc_{24}&=&d_{1,2}^{(24)} + \sum_{j=3}^{n-1} d_{1,j,2}^{(224)} y_j^2,
\end{eqnarray}
\begin{eqnarray}
sc_{42}&=&d_{2,1}^{(24)} + \sum_{j=3}^{n-1} d_{2,j,1}^{(224)} y_j^2,
\end{eqnarray}
\begin{eqnarray}
sc_{26}&=&d_{1,2}^{(26)},
\end{eqnarray}
and
\begin{eqnarray}
\label{eq_sclast}
sc_{62}&=&d_{2,1}^{(26)}.
\end{eqnarray}
Equation~(\ref{eq_scfirst})-(\ref{eq_sclast})
also apply to $f_P$, $f_Q$ and $f_{\Omega}$
if the $d$-coefficients are replaced by the
$p$-, $q$- and $b$-coefficients,
respectively.
The quantities $M_1$ and $M_2$ are defined through
\begin{eqnarray}
M_1
=
-\frac{1}{4} M_{\rm{aux}}
+ \big[
2sc_{00} + (sc_{20}+sc_{02})HR^2
+
\nonumber \\
(sc_{40}+sc_{04})(HR^2)^2
+(sc_{60}+sc_{06})(HR^2)^3
\big]
\zeta
\end{eqnarray}
and
\begin{eqnarray}
M_2=
\zeta^{-1} M_{\rm{aux}} + 
\frac{3}{4} 
\big[
-2sc_{40}-2sc_{04}+2sc_{22}+ \nonumber \\
(-3sc_{60}-3sc_{06}+sc_{24}+sc_{42}) HR^2+\nonumber \\
(sc_{26}+sc_{62})(HR^2)^2
\big] (HR^2)^2 +\nonumber \\
 \big[
-sc_{20}+sc_{02} + (-sc_{40}+sc_{04})HR^2+\nonumber \\
(-sc_{60}+sc_{06})(HR^2)^2
\big]
HR^2 \zeta,
\end{eqnarray}
where 
\begin{eqnarray}
M_{\rm{aux}}=-
\frac{15}{2} (-sc_{26}+sc_{44}-sc_{62}) (H R^2)^4 \zeta^{-1} +\nonumber \\
3 \big[
-sc_{60}+sc_{06}-sc_{24}+sc_{42}
+ \nonumber \\ (-sc_{26}+sc_{62}) HR^2
\big] 
(HR^2)^3.
\end{eqnarray}

\acknowledgments
We thank Javier von Stecher for fruitful discussions
and correspondence.
Support by the ARO and NSF through grant PHY-1205443
is gratefully acknowledged.
This work was additionally supported by the
National Science Foundation through a grant for the Institute for 
Theoretical Atomic, Molecular and Optical Physics at
Harvard University and Smithsonian Astrophysical Observatory.


\begin{thebibliography}{100}
\bibitem{footnotezerominus}
To construct basis functions with $0^-$ symmetry, one
has to couple three spherical harmonics; 
see Refs.~\protect\cite{suzu00,aoya11}.
\end{thebibliography}
\end{document}